%% file: frb_grb_limits.tex
\documentclass[12pt, twocolumn, trackchanges, twocolappendix]{aastex631}

\usepackage{color}
\usepackage{amsmath}
\usepackage{float}
\usepackage{gensymb}
\usepackage{longtable}
\usepackage{threeparttablex}
\usepackage{booktabs}
\usepackage{placeins}

\published{\apj}

\begin{document}
\title{Limits on Fast Radio Burst-like Counterparts to Gamma-ray Bursts using CHIME/FRB}
\shorttitle{}
\shortauthors{}

\input{auth.tex}

\correspondingauthor{Alice P. Curtin}
\email{alice.curtin@mail.mcgill.ca}

\begin{abstract}
Fast Radio Bursts (FRBs) are a class of highly energetic, mostly extragalactic radio transients lasting for $\sim$milliseconds. While over 600 FRBs have been published so far, their origins are presently unclear, with some theories for extragalactic FRBs predicting accompanying high-energy emission. In this work, we use the Canadian Hydrogen Intensity Mapping Experiment (CHIME) Fast Radio Burst (CHIME/FRB) Project to explore whether any FRB-like radio emission coincides in space and time with 81 gamma-ray bursts (GRBs) detected between 2018 July 17 and 2019 July 8 by \textit{Swift}/BAT and \textit{Fermi}/GBM. We do not find any statistically significant coincident pairs within 3$\sigma$ of each other's spatial localization regions and within a time difference of up to one week. In addition to searching for spatial matches between known FRBs and known GRBs, we use CHIME/FRB to constrain FRB-like ($\sim$1-10 ms) radio emission before, at the time of, or after the reported high-energy emission at the position of 39 GRBs. For short gamma-ray bursts (SGRBs), we constrain the radio flux in the 400- to 800-MHz band to be under a few kJy for
$\sim$5.5 to 12.5 hr post-high-energy burst. We use these limits to constrain models that predict FRB-like prompt radio emission after SGRBs. For long gamma-ray bursts (LGRBs), we constrain the radio flux to be under a few kJy from $\sim$6 hours pre-high-energy burst to $\sim$12 hours post-high-energy burst.
\end{abstract}

\section{Introduction} 
\label{sec:intro}
\subsection{FRBs}
Fast radio bursts (FRBs), which were first reported in \citet{lbm+07}, are a class of highly energetic, transient events lasting for a few milliseconds and originating from extragalactic distances. 
While the original discovery consisted of a ``one-off" burst, some FRBs have been shown to repeat \citep{ssh+16a,abb+19b,abb+19c,fab+20}, with two showing periodic burst activity \citep{aab+20, rms+20, Cruces2021}.

FRBs are energetic phenomena, with spectral luminosities $\sim10^{32}$ erg s$^{-1}$ Hz$^{-1}$ over bandwidths of $\sim0.1$ to 1 GHz. While many models for FRB progenitors involve neutron stars (NS), these NS-based models either require substantial scaling up of previous models for NS emission mechanisms, e.g., curvature radiation in \citet{klb17} or synchrotron radiation in \citet{2011Sironi}, or require new astrophysical mechanisms to provide the necessary energy. Proposed new mechanisms include interactions of magnetized relativistic shockwaves \citep{mms19}, a stream of particles combing through the magnetosphere of the NS \citep{zhang2017}, an asteroid belt surrounding a pulsar \citep{dwwh16}, or an expanding fireball \citep{Ioka2020}. In addition to radio emission, these mechanisms might also produce accompanying, detectable high-energy emission \citep[e.g.,][]{zhang2017,mms19} as the radio emission is expected to be a small fraction of the total energy output. 
\subsection{High-energy Emission from FRBs}
The earliest observational evidence for accompanying high-energy emission for an FRB came when \citet{2012Bannister} detected single radio pulses after two gamma-ray bursts (GRBs) using the 12-m dish at the Parkes Observatory. However, they were unable to rule out radio frequency interference (RFI) as the cause of these pulses, and their simulations/modeling predicted that there was only a 2$\%$ chance that the pulses were actually associated with the GRBs. Additionally, \citet{2016Delaunay} claimed an association between FRB 20131104A and an untriggered GRB found in archival \textit{Swift}/Burst Alert Telescope (BAT) data. However, a more recent analysis done by \citet{2021Sakamoto} of the same \textit{Swift}/BAT dataset found no significant gamma-ray emission at the time of FRB 20131104A.

More conclusive observational evidence for the association of FRBs with high-energy emission came in April 2020 when the Canadian Hydrogen Intensity Mapping Experiment Fast Radio Burst (CHIME/FRB) Project detected an ultra-bright FRB-like pulse from the Galactic magnetar SGR 1935+2154 \citep{2020CHIMESGR}. The burst was detected at radio frequencies by CHIME/FRB and STARE2 \citep{2020NatureBochenek}, and in the hard X-ray/soft gamma-ray regimes by the Hard X-ray Modulation Telescope (\textit{Insight}-HXMT) Satellite \citep{2021NatureInsightSGR}, the Konus-\textit{Wind} Experiment \citep{2020ATelKonusWindSGR}, and the INTErnational Gamma-Ray Astrophysics Laboratory \citep[\textit{INTEGRAL},][]{2020IntegralSGR}. The burst detected at CHIME/FRB had a radio-to-gamma-ray (20 to 200 keV) fluence ratio of  $9^{+9}_{-5} \times 10^{11}$ Jy ms erg$^{-1}$ cm$^2$ ($4^{+4}_{-1.8} \times 10^{-6}$ in dimensionless units assuming a 400-MHz bandwidth), with the dispersion-corrected arrival times of the subcomponents of the burst at CHIME/FRB leading the high-energy emission at Insight-\textit{HXMT} by $\sim$3 ms \citep{2020IntegralSGR}. 

Numerous simultaneous X-ray and high-energy observations, as well as follow-up observations and searches through archival data, have been carried out to find detected high-energy associations with an extragalactic FRB \citep[e.g.,][]{sbh+17, Cunningham2019,Martone2019, Cunningham2019, scholz2020, Guidorzi2020, 2020SwiftGuano, Casentini2020, Anumarlapudi2020, Verrecchia2021, Principe2021arXiv, Mereghetti2021}. Despite these many efforts, no other unambiguous detections have been made. 


Possibilities for the aforementioned lack of detections include limited sensitivity of high-energy instruments, observations which do not overlap with FRB bursts \citep[e.g.,][]{Mereghetti2021, Guidorzi2020} and different beaming angles for the radio and high-energy emission \citep[e.g.,][]{Sridhar2021Wind}. One way of overcoming some of these challenges is to search for radio bursts associated with high-energy emission, rather than searching for high-energy bursts associated with radio emission. For the remainder of this paper, we specifically focus on searching for FRB-like radio emission from GRBs. 

\subsection{GRBs}GRBs are extremely energetic, extragalactic sources of gamma-ray emission with luminosities $>10^{53}$ erg s$^{-1}$ \citep{1973GRBDetection, 2009Gehrels}. Since their initial discovery in 1967, over 7000 GRBs have been detected, with GRBs now classified as either short GRBs (SGRBs), whose 90$\%$ burst duration (T90) is less than 2 s, or long GRBs (LGRBs), whose T90 is longer than 2 s. However, making an unambiguous distinction between the two classes requires more than just duration as SGRBs typically also have a harder spectrum than LGRBs \citep[][]{1993Kouveliotou, 2009Ghirlanda, 2020Jespersen}.

Most SGRBs are likely produced by compact object mergers such as black hole (BH)-NS mergers or NS-NS mergers \citep{2014Berger}. Some nearby (e.g., less than a few tens of Mpc) SGRBs can be produced by magnetar giant flares though \citep[][]{2021ApJ...907L..28B, 2021NatAs...5..385F, 2021Natur.589..211S}. However, since SGRBs produced by magnetar giant flares are a very small fraction of the observable SGRB population due to the limited distance range to which we can detect them, we do not focus on them in detail when discussing models for FRB-like emission from GRBs, and instead only focus on SGRBs produced by NS-NS or NS-BH mergers.

LGRBs are traditionally thought to be produced during the collapse of a rapidly rotating massive star \citep{2019GalYam}. However, a LGRB was recently associated with a kilonova, suggesting that some LGRBs can similarly be produced by compact object mergers \citep{2022LGRBKilonova}.

\subsection{Theoretical Predictions for Radio Emission from GRBs}
Assuming that SGRBs originate from the merger of compact objects, then an FRB-like burst might be produced prior to the SGRB from precursor winds \citep{Lyutikov2013, Sridhar2021Wind}, interactions between a highly magnetized NS and a less magnetized NS under a unipolar inductor model \citep{2012Piro, Wang2016}, synchronization of two NS's magnetic fields to the binary orbital period \citep{Totani2013}, magnetic flares produced by the twisting of magnetospheres \citep{MostPhilippov2020}, an induced electric field due to the motion of a magnetized NS \citep{Wada2020}, or  abrupt magnetic reconnections due to interactions between two NS's magnetospheres \citep{Zhang2020}. Pre-merger emission is also possible for BH-NS binaries, e.g., from a BH-NS battery \citep{Mingarelli2015}, from an induced electric field due to the motion of a magnetized NS around a BH \citep{Wada2020}, or from extraction of orbital kinetic energy due to the NS orbiting the BH \citep{Carrasco2021}. Emission at the approximate time of the SGRB is also possible from interactions between a highly magnetized wind produced by the merger and the surrounding medium \citep{UsovKatz2000}. However, although this emission would be produced approximately simultaneously, its detection would be significantly delayed from the time of the high-energy emission due to the intergalactic dispersion of radio waves. This intergalactic dispersion is quantified using the dispersion measure (DM) delay of the radio emission, and allows rapid radio follow-up of GRBs to possibly detect radio emission that was emitted simultaneously with the high-energy emission. 

FRB-like emission may also be possible after a NS-NS or NS-BH merger. If a NS-NS merger results in another NS, then given certain NS parameters (e.g., high magnetic field strength, fast spin period) pulsar-like emission from the NS could be on the same energy scale as that of FRBs \citep{ PshirkovPostnov2010,Totani2013}. If the NS is stable (dependent on its mass, spin, spin-down, and the NS equation of state), then this emission could theoretically last indefinitely. However, if the NS is unstable and collapses to form a BH, then an additional FRB-like burst might be produced from the transfer of the NS's magnetic field to the BH \citep{FalckeRezzolla2014, zhang2014}. In a NS-BH merger, an FRB-like burst might be produced in a similar manner when the NS's magnetic field is transferred to the BH \citep{Mingarelli2015}.

There are fewer theoretical predictions for prompt FRB-like radio emission from LGRBs. The host environment of the first detected repeating FRB, FRB 20121102A, is very similar to that of LGRBs as it is a star-forming dwarf galaxy with low metallicity \citep{tbc+17}. This provides some evidence for an association between FRBs and LGRBs \citep{mbm17}. However, FRBs have since been found in other host galaxy environments \citep[e.g.,][]{2021Niu, 2021Bhardwaj, 2021Kristen, 2022Bhandari}, and no definitive association has been made between an LGRB site and an FRB site. Additionally, while the gamma-ray emission after a supernova could escape, the surrounding region would likely be opaque to radio emission on the timescale of approximately years to decades, and thus not simultaneously detectable. It is possible that if an LGRB were produced in a unconventional mechanism, e.g., a merger-driven explosion induced by a compact object as suggested by \citet{Dong2021}, then there might be the possibility of pre-LGRB radio emission, but this has yet to be theoretically explored. Additionally, a recent detection of an LGRB associated with a kilonova suggests that LGRBs can be  produced by compact object mergers, and hence some of the above mentioned mechanisms for producing radio emission at the time of SGRBs could be applied to some LGRBs  \citep[][]{2022LGRBKilonova, 2022YangLGRBMerger}.

\subsection{Observational Efforts to Search for Radio Emission from GRBs}
There have already been a number of searches for FRB-like radio emission associated with SGRBs and LGRBs, but none has been successful\footnote{During the preparation of this manuscript, a tentative association was claimed between FRB 20190425A and the binary neutron star merger GW 20190425 \citep{2022arXiv221200201M}. There is one candidate SGRB that may be related to this event, but its significance is low and its uncertainty region high \citep{2020AstL...45..710P}. No definitive association has been made between the FRB and this candidate SGRB. }. \citet{2012Bannister}, \citet{2013Staley}, \citet{Palaniswamy2014}, \citet{Kaplan2015}, \citet{2021Anderson}, \cite{2021Rowlinson} all performed rapid radio follow-up of GRB locations, while \citet{Madison2019}, \citet{Men2019}, \citet{2020Hilmarsson}, and  \citet{2021Bruni} performed late-time follow-up of GRB sites. Some of the GRBs followed-up by \citet{Madison2019} and \citet{Men2019} were notable for possibly being associated with NS remnants due to anomalous X-ray activity and their X-ray light curves \citep{2014Fong, 2014MetzgerPiro, 2017Lu, 2019Troja, 2019Lamb}. However, no FRB-like radio emission was detected in any of the above searches. In addition to the above follow-up, \citet{Obenberger2014} performed radio observations that were simultaneous with high-energy detections, yet no radio emission was seen. \citet{tkp16} also performed radio observations of the SGR 1806--20 magnetar giant flare (relevant as some magnetar giant flares produce SGRBs), but again no FRB-like radio burst was detected.

 It is possible that the sources followed up in the above studies either do not produce FRBs or produced FRBs at a time different than that of the follow-up observations. To further explore whether FRBs and GRBs might be related, we take advantage of the unique wide field of view and large exposure of CHIME/FRB and search for radio emission before, at the time of, and after 81 GRBs. Below, in Section \ref{section: chime overview}, we provide a basic overview of the CHIME telescope and the CHIME/FRB backend. In Section \ref{sec: sources}, we describe our GRB and FRB samples. Then, in Section \ref{sec: chance coincidence}, we outline our search for temporally and spatially coincident GRBs and FRBs and present our results. In Section \ref{sec: flux limits}, we outline our pipeline for calculating radio upper limits at times of non-detections and discuss the results from this pipeline. In Section \ref{sec: discussion}, we discuss the implications of our results for different SGRB and LGRB models. We end in Section \ref{sec: summary} by summarizing our work.

\section{Overview of CHIME/FRB} \label{section: chime overview}
CHIME consists of four, $100$-m $\times$ 20-m, cylindrical, parabolic reflectors with the cylinder axes oriented in the north-south (N-S) direction. It has a N-S field of view (FOV) of $\sim120\degree$ and an east-west (E-W) FOV of $\sim1.3$ to 2.5$\degree$ (dependent on frequency, with 1.3$\degree$ corresponding to 800-MHz) for a total FOV of $\sim250\degree$ \citep{nvp+17}. Each of the four reflectors has 256 dual-polarization feeds hanging along its axis for a total of 2048 antenna signals operating between 400- and 800-MHz. 

The 2048 signals are fed into a hybrid FX correlator where the F-engine digitizes and transforms the antenna signals and the X-engine performs spatial correlation. More specifically, the X-engine forms 256 N-S beams through a spatial fast Fourier transform (FFT) of the antenna signals, and then forms three more sets of 256 N-S beams using exact phasing for a total of 1024 stationary formed beams. For more details on the formed beams, see \citet{nvp+17} and \citet{msn+19}. 

After the X-engine, the data are sent to the CHIME/FRB pipeline, which is split into four processing levels: L1, L2, L3, and L4. First, L1 performs RFI mitigation \citep{2022MasoudRFI}, performs a fast dedispersion transform, and identifies the candidates with FRB-like signals.
Then, L2/L3 determine whether a candidate is RFI, a known source, or an unknown source, and decide which set of actions to perform for the candidate based on that classification.
L4 then implements the actions that are decided in L2/L3. For more details on CHIME/FRB and its pipelines, see \citet{abb+18}. More details on CHIME and its cosmology experiment can be found in  \cite{2022chimeoverview}.

\section{Sources} \label{sec: sources}

\subsection{FRBs} \label{sec: FRB sample}
The FRB sample used in this analysis consists of 536 FRBs detected between 2018 July 25 and 2019 July 1 from the first CHIME/FRB catalog. This includes 62 bursts from 18 previously reported repeating FRBs, and 474 thus far one-off FRBs. A detailed analysis and description of these FRBs can be found in  \citet{chimefrbcatalog}.

\subsection{GRBs} \label{sec: GRB sample}

The GRB sample used in this analysis consists of 81 GRBs detected between 2018 July 17 and 2019 July 8. The time-frame for these GRBs is chosen to be the same as that of the first CHIME/FRB Catalog. However, we also include GRBs from 7 days before and 7 days after the time of the catalog, with the reasoning for this discussed below in Section \ref{sec: chance coincidence}. The GRBs are chosen using the Gamma-ray Coordination Network (GCN)\footnote{www.gcn.gsfc.nasa.gov} circulars from four different high-energy instruments: the \textit{Fermi} Gamma-ray Burst Monitor, \citep[\textit{Fermi}/GBM;][]{Meegan2009}, the \textit{Neil Gehrels Swift Observatory} Burst Alert Telescope \citep[\textit{Swift}/BAT,][]{Gehrels2004, Barthelmy2005}, \textit{INTEGRAL} \citep[][]{Winkler2003}, and Konus-\textit{Wind} \citep{Aptekar1995}. 
We also restrict our analysis to GRBs having 1$\sigma$ localization errors in RA and DEC less than $0.5\degree$ as it is not possible to make conclusive statements on spatial coincidences or calculate radio flux limits for GRBs with either unknown or large uncertainty regions.  

Many of the GCN circulars contain false positives, and we manually remove all of these from our sample along with all bursts for which either the T90 or flux/fluence could not be determined. The high-energy fluxes and fluences, along with the T90s, were not readily available in the GCNs, and were instead determined using the respective online GRB catalogs \citep{2016Lien, 2014Gruber, 2014VonKienlin, 2016Narayana, 2020VonKienlin}. To best differentiate SGRBs from LGRBs, we would ideally have both the T90 and the spectral hardness of each burst. However, in this work, we limit ourselves to classifications based solely on the T90s of the bursts, with GRBs whose T90 is less than 2 s classified as SGRBs, and those whose T90 is greater than 2 s classified as LGRBs. We do manually compare our classifications with those found online though. The majority of the GRBs we classify as LGRBs do not have available online classifications, but their durations are $\gg2$ s. Of the SGRBs in our sample, two SGRBs, GRB 190326A and GRB 190627A, are unique in that while they are short, they are also spectrally soft, and SGRBs are expected to be spectrally hard \citep{2019GCN.24020....1M, 2019GCN.24899....1B}. Thus, while we still classify them as SGRBs, we note that they may be exceptions. Additionally, while GRB 190331A's T90 is 4.03 s, it is noted to likely be a SGRB in a GCN Notice.\footnote{https://www.mpe.mpg.de/~jcg/grb190331A.html} Thus, we include GRB 190331A in our SGRB sample. 


\section{Searching for Temporally and Spatially Coincident FRBs and GRBs} \label{sec: chance coincidence}

\subsection{Search}\label{sec: search pairs}

Our primary interest is searching for GRBs and FRBs that are spatially and temporally coincident. To search for spatial coincidence between a GRB and FRB, we look for FRB and GRB pairs where the GRB's 3$\sigma$ localization overlaps with part of the 3$\sigma$ localization region of the FRB. Note that the average 1$\sigma$ localization radius of our sample of GRBs is 0.04$\degree$ and the average 1$\sigma$ localization radius of our sample of FRBs is 0.27$\degree$. The localization regions used in this analysis for our FRB sample are determined using CHIME/FRB Metadata headers \citep[for more details on CHIME/FRB Metadata headers, see][]{chimefrbcatalog}. For the 3$\sigma$ localization regions for both the GRBs and FRBs, we assume $3 \times 1\sigma$ localization. For our FRBs, we refer to this localization radius as the 3$\sigma$ CHIME/FRB localization uncertainty. However, this assumes that the localization regions are Gaussian, which is not true for CHIME/FRB FRBs (see \citet{chimefrbcatalog} for more details on this). The true localization regions are, however, highly complex multi-region geometries \citep[e.g., see Figure 6 in][]{chimefrbcatalog} and non-trivial to search over. Thus, rather than searching over the true localization region for each FRB, we instead perform an additional, more conservative search in which we re-do our joint temporal and spatial coincidence search using a radius of 1$\degree$ for our localization regions for all our FRBs (and hence a 3$\sigma$ radius of 3$\degree$). Then, for any coincidences when using this method, we manually check the coincidences using the true CHIME/FRB localization regions presented in \citet{chimefrbcatalog}. As a reference, the FWHM of the CHIME/FRB primary beam used in this work is $\sim$1.6$\degree$ at 400-MHz, and the FWTM is $\sim$3$\degree$ at 400-MHz. Thus, searching within a 3$\degree$ radius is equivalent to searching within a radius equal to the FWTM of the primary beam. 

\begin{figure}[]
    \centering
    \includegraphics[width=0.9\linewidth] {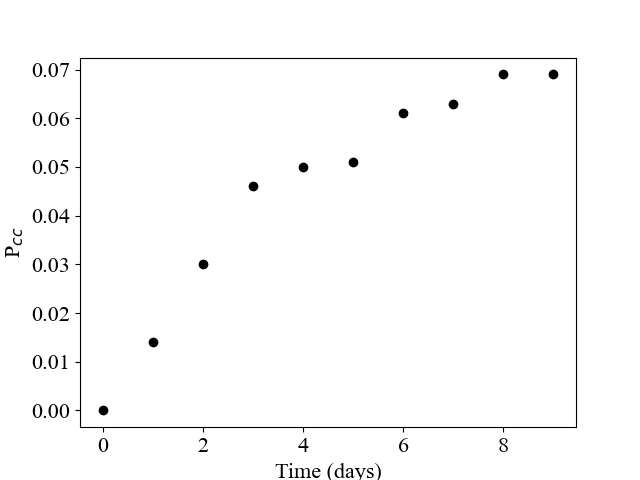}
    \caption{Probability of having at least one temporal and spatial chance coincidence (P$_{cc}$) as a function of the time difference in days between a given GRB and FRB. Simulation is performed using our true sample of GRBs and a simulated sample of FRBs. We assume a 3$\sigma$ localization radius of 0.81$\degree$ for our simulated FRBs.}
    \label{fig:Pcc fnct time}
\end{figure}

Temporally, we search for FRBs and GRBs that are coincident within periods up to seven days. While the upper limit of seven days is chosen slightly arbitrarily, we are primarily interested in coincident high-energy and radio emission and hence a small time window around the high-energy emission. Additionally, as shown in Fig. \ref{fig:Pcc fnct time}, the probability of having at least one chance coincidence association for periods greater than 7 days is $\gtrapprox5\%$ (the method for calculating this is discussed in detail below in Section \ref{subsec: cc calculations}) and thus it would be difficult to conclude that the association is significant. The search of $\pm$ 7 days is the reason for the GRB sample window extending from 2018 July 17 to 2019 July 8 rather than from 2018 July 25 to 2019 July 1.

When applying the joint temporal (up to seven days) and spatial criteria (overlapping 3$\sigma$ CHIME/FRB localization radius), there are no GRBs within our sample that are coincident with any of the FRBs from the first CHIME/FRB catalog. The same result holds when using a 3$\sigma$ localization radius of 3$\degree$ for the FRBs. However, when considering solely spatial coincidence, and using the 3$\sigma$  CHIME/FRB localization uncertainties for our FRBs, there are two GRBs in our sample that are consistent with two FRBs from the first CHIME/FRB Catalog. These GRBs are GRB 180914B and GRB 181231A with corresponding FRBs FRB 20190614B and FRB 20190109B, respectively. The time differences, $\textrm{t}_\textrm{FRB} - \textrm{t}_\textrm{GRB}$, for the events are 273 days and 10 days, respectively, and the spatial differences are $0.45\degree$ and $1.29\degree$ between the most likely positions. The total quadrature summed 1$\sigma$ uncertainties on the spatial positions for the two pairs are $0.33\degree$ and $0.46\degree$, respectively. 

\subsection{Calculating Chance Probability of FRB-GRB Associations} \label{subsec: cc calculations}

To determine whether or not these two spatial coincidences are astrophysically significant, we perform two different tests of the chance probability of two coincident FRB-GRB pairs within our sample. First, we simulate a set of 81 GRBs uniformly distributed on the sphere. We also randomly distribute these simulated GRBs over the time interval covered by the first CHIME/FRB catalog. For the spatial error on this set of GRBs, we use the average 3$\sigma$ spatial error from our true sample of 81 GRBs, 0.12\degree. We then cross-check the spatial positions of this sample of simulated GRBs with our real sample of 536 FRBs. For our real sample of FRBs, we use the 3$\sigma$ CHIME/FRB localization uncertainties. Performing 1000 different simulations, we find there is a 41\% chance of detecting two or more pairs of spatially coincident GRBs and FRBs within our given samples.

\begin{figure}[]
    \centering
    \includegraphics[width=0.9\linewidth]{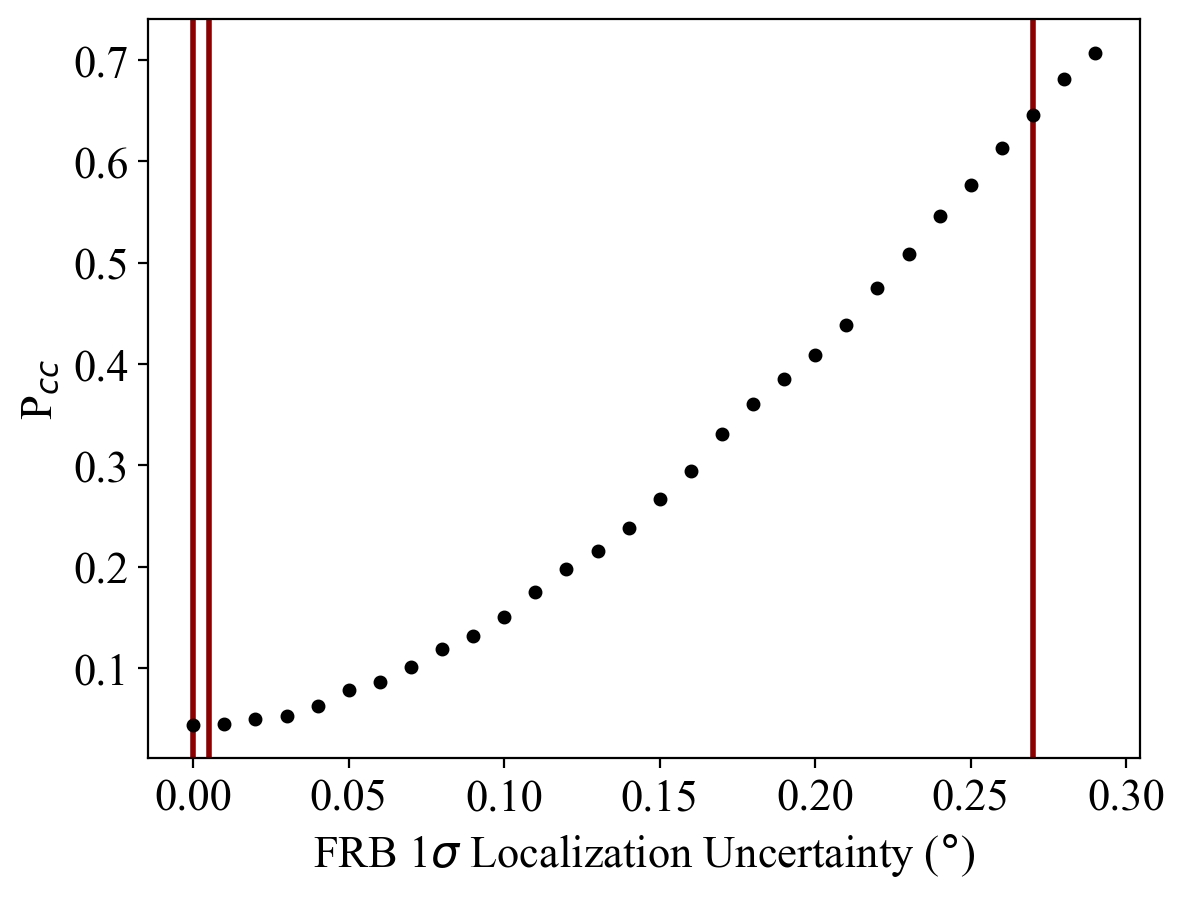}
    \caption{Probability of having at least two spatial chance coincidences (P$_{cc}$) as a function of the 1$\sigma$ uncertainty on the FRB's position. Simulation is performed using our true sample of GRBs for which the average 3$\sigma$ uncertainty is $0.12\degree$, and a simulated sample of FRBs. The leftmost red line indicates the average positional uncertainty of the future CHIME/Outriggers Experiment \citep[$\sim$ 50 mas;][]{2022Cassanelli}, the second leftmost red line indicates the average positional uncertainty of the CHIME/FRB baseband system \citep[$<1$ arcmin;][]{2021DanielleBaseband}, and the rightmost red line indicates the average positional uncertainty of the CHIME/FRB header localizations ($\sim0.27\degree$).}
    \label{fig:Pcc fnct sep}
\end{figure}

For our second test, instead of simulating our GRBs, we simulate a set of 536 FRBs. Using the first CHIME/FRB catalog, we generate a 2D probability distribution in declination and time. We independently sample from the declination and time axis in order to produce a set of declinations and timestamps, and use our timestamps to determine a distribution of simulated RAs within $\sim$1.6$\degree$ of the meridian as this is the FWHM of the primary beam at 400-MHz. We combine the sampled declinations with these RAs to produce a simulated set of FRBs. For the spatial error on each FRB, we use a 3$\sigma$ localization radius of 0.81$\degree$. We then cross-check the spatial positions of this sample of 536 simulated FRBs with our true sample of 81 GRBs. Performing 1000 simulations, we find there is a 65$\%$ chance of detecting two or more spatially coincident FRBs and GRBs within our given samples. In Figure \ref{fig:Pcc fnct sep}, we show the chance probability of having at least two spatial FRB-GRB pairs as a function of the 1$\sigma$ FRB uncertainty. Our 1$\sigma$ localization uncertainties would need to be $\ll1\degree$ in order for an association to be significant.  

This method of simulating a set of FRBs and comparing it with the known sample of GRBs is also used to determine the chance probability of having both a temporal and spatial association as a function of the time differences (e.g., $|\textrm{t}_\textrm{FRB} - \textrm{t}_\textrm{GRB}|$) as discussed earlier in Section \ref{sec: search pairs} and shown in Figure \ref{fig:Pcc fnct time}.

Both of these methods argue that the two spatially-coincident pairs found are not astrophysically significant. We therefore conclude that there are no significant detections of coincident FRBs and GRBs within our sample. 

\section{Radio Upper Limits} \label{sec: flux limits}
\subsection{Determining Radio Upper Limits}  \label{subsec: calculating radio upper limits}
In addition to searching for coincident FRB and GRB pairs, we also place limits on the possible 400- to 800-MHz radio flux for GRBs which are within the FOV of CHIME/FRB at a point either before, at the time of, or after their high-energy emission. There are seven GRBs that are within the FOV of CHIME/FRB at the time of their high-energy emission, and another 32 that are within the FOV either six hours before or 12 hours after the high-energy emission (these time windows are discussed in detail in Section \ref{sec: results}). These 39 GRBs are also detected on nominal operational days, and have radio flux limits (discussed in detail below) $<100$ kJy. Here, we limit the FOV to within $\sim$17$\degree$ of CHIME's meridian, as the CHIME beam is well-modeled within this range, and it is outside the scope of this work to model the beam at locations much further from meridian \citep{chimefrbcatalog, 2022dallasprimarybeam, 2022chimeoverview}. We calculate an upper limit on the FRB-like radio emission by combining our beam sensitivity at the location of the GRB with the flux to signal-to-noise ratio (S/N) of a spatially nearby calibrator FRB. 

More specifically, to calculate an upper limit on radio emission at a specific time and sky position, we would ideally have a near-simultaneous and near-on-the-sky source for our flux calibration. As described in \citet{abb+18}, CHIME/FRB performs flux calibration using steady sources. These sources are used to determine a conversion between the beamformed intensity units (e.g., the output units after the FFT beamforming) and the conventional flux units of Jy. However, this process is not dependent on, nor does it account for, the CHIME/FRB detection S/N. As our goal is to convert a given minimum detection S/N to a flux or fluence, a new calibration process is needed for this work.

Instead of using steady-sources, we use previously detected and calibrated FRBs as our GRB-flux calibrators. Ideally, these FRBs would be near-simultaneous and near-on-the-sky to the GRB. However, as CHIME/FRB only detects $\sim2$ to 3 FRBs per day, it is not possible to satisfy both criteria. Instead, we choose an FRB nearby in declination (e.g., $<1\degree$ away\footnote{This angular separation is equivalent to the width of $\sim$2-3 N-S CHIME/FRB beams.}) to a given GRB and use models of our beam shape to account for slight differences in position between the two sources \citep{nvp+17, msn+19, chimefrbcatalog, 2022chimeoverview}. 

We do not restrict the timestamp of the calibrator FRB and instead use system sensitivity metrics to account for temporal differences by scaling our radio limits by the ratio of the system sensitivity at the time of the FRB with that at the time of the GRB. These system sensitivity metrics are determined using the daily root mean square (RMS) noise of pulsar detections with CHIME/FRB, with a detailed description of these metrics provided by \citet{jcf+19} and \citet{fab+20}. We do not account for fluctuations in the system sensitivity on a timescale smaller than one day as these metrics were not consistently available. 
However, as discussed by \citet{Josephy2021}, bandwidth effects due to RFI and correlator cluster health typically only result in average sensitivity losses of $\sim$10$\%$. While CHIME/FRB has developed a parallel injections system \citep[][]{2022Merryfield} to quantify further aspects of the system's sensitivity to given parameters, this has not been applied in this work. However, as discussed by \citet{chimefrbcatalog}, the only known major bias to-date of the CHIME/FRB pipeline is against highly scattered bursts. This does not present a large bias against redshift though, as the two main contributors to the observed scattering timescales of FRBs are the circumburst environment and intervening galaxies/galaxy halos \citep{Chawla2021}, and scattering in both locations is inversely proportional to redshift \citep{2022CordesOckerChatterjee}.

There is an additional, unaccounted DM delay between any detection of high-energy emission and the time at which a simultaneous radio burst would be detected by CHIME/FRB. Such a DM delay can, in principle, be estimated if the distance or redshift of the GRB is known. Unfortunately, only eight of our 81 GRBs have readily available redshifts. For the GRBs without redshifts, we try three different redshift/DM estimates. First, we use the median redshift of the respective SGRB or LGRB redshift distribution as an approximate redshift for each GRB ($z=0.82$ for SGRBs and $z=1.71$ for LGRBs). The distributions are created using the online GRB catalog GRBWeb \citep{GRBweb}. Second, we use the minimum redshift from the SGRB and LGRB redshift distributions (0.038 for SGRBs and and 0.0085 for LGRBs).
Lastly, instead of using a redshift, we use 1.2$\times$ the maximum DM detected in the first CHIME/FRB catalog (3645 pc cm$^{-3}$). The extra 20\% was chosen arbitrarily to allow for an event detected further out into the tail of the DM distribution than CHIME/FRB has sampled thus far.

For cases in which we have a redshift rather than a DM, e.g., either the true redshift or an approximate redshift, we convert the redshift to a DM by assuming:
\begin{equation}
    \textrm{DM}_{\textrm{tot}} = \textrm{DM}_{\textrm{MWdisk}} + \textrm{DM}_{\textrm{MWhalo}} + \textrm{DM}_{\textrm{IGM}} + \textrm{DM}_{\textrm{Host}}
\end{equation} 
where $\textrm{DM}_{\textrm{MWdisk}}$ is the Milky Way disk (MW) contribution, $\textrm{DM}_{\textrm{MWhalo}}$ is the MW halo contribution, $\textrm{DM}_{\textrm{IGM}}$ is the intergalactic medium contribution (IGM), and $\textrm{DM}_{\textrm{Host}}$ is the contribution from the host galaxy. We calculate $\textrm{DM}_{\textrm{MW}}$ for each individual GRB by using the average of the NE2001 \citep{cl02} and the YMW16 \citep{ymw17} electron density models along the respective LOS. We assume a $\textrm{MW}_{\textrm{Halo}}$ DM of 50 pc cm$^{-3}$ \citep{pz19}, estimate $\textrm{DM}_{\textrm{IGM}}$ using the Macquart relation \citep{mpm+20}, and assume a host galaxy DM of 175 pc cm$^{-3}$/(1+z) as suggested by \citet{Chawla2021}. By using the minimum of the redshift distribution and 1.2$\times$ the maximum DM from the first CHIME/FRB catalog in addition to the median redshift, our DM ranges should cover uncertainties in the host galaxy DM and MW halo DM. 

While $\textrm{DM}_{\textrm{tot}}$ depends on the Galactic longitude, Galactic latitude, and redshift of the GRB, the average $\textrm{DM}_{\textrm{tot}}$ when using the median redshift of the respective distribution is $\sim$1142 pc cm$^{-3}$ for SGRBs and $\sim$1845 pc cm$^{-3}$ for LGRBs corresponding to delays of $\sim$29 s and $\sim$47 s at 400-MHz, respectively. The average $\textrm{DM}_{\textrm{tot}}$ when using the minimum redshifts are $\sim$300 pc cm$^{-3}$ for SGRBs and $\sim$400 pc cm$^{-3}$ for LGRBs. The upper limits we present are the most conservative limits associated with the timestamps from the three possible DM delays. Note that in the vast majority of cases ($>$95$\%$), the upper limits calculated using either the minimum redshift or maximum DM are still within the 3$\sigma$ uncertainty of the upper limit calculated using the median redshift. 

Due to the differences in position between our calibrator FRB and the GRB of interest, we must scale the flux of the FRB by differences in CHIME/FRB's beam sensitivity at the two positions. CHIME/FRB's spatial sensitivity can be described by a combination of two different beam models: that of the individual antennas which we will refer to as the primary beam model \citep[discussed in detail by][]{chimefrbcatalog, 2022chimeoverview}, and that of the interferometric formed beams \citep[discussed in ][]{nvp+17}. The beam response of CHIME/FRB in the far sidelobes is still being characterized, and thus an FRB detected in the sidelobes of CHIME/FRB could both have an inaccurately determined position and a significantly underestimated flux. In order to avoid using FRBs detected in the sidelobes as our calibrator FRBs, we impose the condition that the bandwidth of our calibrator FRBs be sufficiently large (e.g., larger than $3/4$ of our 400-MHz bandwidth), as the bandwidth of an FRB detected in the sidelobe would be significantly modulated (Hsiu-Hsien et al. \textit{in prep.}).

We use our calibrator FRB to scale the FRB's measured flux/fluence \citep[see][for details on the flux/fluence measurements from the first CHIME/FRB catalog]{chimefrbcatalog} to detection S/N. However, as described in detail by \citet{chimefrbcatalog}, the flux and fluence for each FRB is calculated assuming the FRB is located on the meridian. Thus, the flux/fluence is the estimated flux/fluence for a given S/N burst on the meridian. Rather than correcting the flux/fluence to S/N for the true position of the FRB, and then scaling it to the location of the GRB, we instead directly scale the flux/fluence to S/N of the calibrator FRB by the ratio of the primary beam sensitivity at meridian with the primary beam sensitivity at the location of the GRB. Additionally, we must scale this ratio by the interferometric formed beam sensitivity at the location of the GRB. The published fluxes and fluences for FRBs also assume the FRBs are at the centers of the formed beams, so we take the formed beam sensitivity for the FRB to be unity.

As both the primary and formed beams are frequency dependent, we must make assumptions on the spectral shape of the emission for the calibrator FRB and the GRB. As in the first CHIME/FRB catalog \citep{chimefrbcatalog}, we assume a flat spectral index for all of our calibrator FRBs. We similarly assume a flat spectrum for all of our GRBs, as there is not yet any observational evidence to guide this. We note that the spectral indices used in the emission models presented by \citet{RowlinsonAnderson2019} all assume a negative spectral index, although the magnitude of the spectral index is not well constrained in any model. Additionally, while our FRB sample has measured bandwidths\footnote{We do not correct these bandwidths for instrumental effects in this work.}, we assume a bandwidth of 400-MHz for the GRBs. This assumption affects our estimates of our beam sensitivity, as these sensitivities are highly frequency dependent. 

Assuming conservatively that CHIME/FRB is sensitive to bursts with a S/N threshold of 10, we scale the flux/fluence to detection S/N of our calibrator FRB to a S/N of 10. Combining the calibrator FRB's flux to S/N ratio, the system sensitivities, and the beam sensitivities, the upper limit on the radio flux in the 400- to 800-MHz CHIME/FRB band is given by:

\begin{equation}
   \textrm{Flux}_{\textrm{GRB}} = 10 \times  \frac{\textrm{Flux}_{\textrm{FRB}}}{\textrm{S/N}_{\textrm{FRB}}} \times  \frac{\textrm{B}_{\textrm{M}}}{\textrm{B}_{\textrm{GRB}}} \times 
    \frac{1}{\textrm{F}_{\textrm{GRB}}} \times 
    \frac{\Delta\textrm{S}_{\textrm{Sys,GRB}}}{\Delta\textrm{S}_{\textrm{Sys,FRB}}}
\label{eq: fluxUpperLimit}
\end{equation}

\normalsize 
\noindent where $\textrm{B}$ is the primary beam response, $\textrm{F}$ is the formed beam response, $\Delta\textrm{S}_{\textrm{Sys}}$ is the system sensitivity\footnote{The system sensitivity is determined by RMS noise, so a higher value for the system sensitivity means we were less sensitive on that day, thus the ratio is flipped for this one quantity.}, and $\textrm{M}$ stands for meridian. A similar equation can be applied for an upper limit on the fluence of a radio burst from a GRB. While not shown in Eq.~\eqref{eq: fluxUpperLimit}, we scale our limits to an assumed radio burst width of 10-ms. Additionally, for all GRBs (or FRB calibrators) within 10$\degree$ of the Galactic plane, we scale Eq. ~\eqref{eq: fluxUpperLimit} by the ratio of the sky temperature at the two locations using the 2014 Haslam all-sky continuum map at 408 MHz \citep{2015Haslam}. We assume an observing frequency of 600-MHz for CHIME/FRB and a brightness temperature spectral index of $-2.5$ for the Galactic synchrotron emission when calculating the sky temperatures \citep{2003Bennett}.

The uncertainties on the upper limits are dominated by the uncertainties of the fluxes of our FRB calibrators, but are also determined using the uncertainties on our primary beam model, our system sensitivity, and the GRB localization. To quantify the effect of the GRB localization uncertainties on our final flux and fluence upper limits, we run a Monte Carlo (MC) in which for every timestamp for a given GRB, we re-calculate the formed beam and primary beam sensitivities at 1000 different locations within the 3$\sigma$ uncertainty region of the GRB. We quantify the effect that this MC has on our final quoted upper limits by calculating the difference in the GRB's flux density limit uncertainty before and after this MC step as compared with the GRB's flux density limit e.g., $|\sigma_{\textrm{initial}} - \sigma_{\textrm{final}}|/ \textrm{flux}_{\textrm{GRB}}$. In all but two cases (discussed below) this difference is negligible e.g., $|\sigma_{\textrm{initial}} - \sigma_{\textrm{final}}| / \textrm{flux}_{\textrm{GRB}}< 0.05$. This is expected since the average 3$\sigma$ localization uncertainty of our GRB sample is small e.g., 0.12$\degree$. The two cases for which the above does not hold are  GRB 181120A which has the largest 1$\sigma$ positional error of our sample of 0.5$\degree$ and GRB 180914B which has the second largest 1$\sigma$ positional error of 0.16$\degree$. Thus, for these two GRBs, we include an additional MC step when determining the flux and fluence upper limit uncertainties.


\subsection{Results} \label{sec: results}
While we do not find any GRBs that are spatially and temporally coincident with our FRBs, ten GRBs are within the FOV of CHIME at the time of their high-energy detection (after accounting for the additional dispersion delay). However, two of these high-energy bursts occurred during periods of non-nominal telescope operations (GRB 181002A and GRB 180930A), and are removed from this analysis. A third burst is located at the edge of CHIME's FOV (GRB 190326A), and is removed due to an estimated flux limit $>$100 kJy due to the lack of constraining power from such a high flux density limit. In addition to the remaining seven bursts, there are 32 bursts which are within the FOV (assumed to be within $\sim$17$\degree$ of CHIME's meridian as discussed above in Section \ref{subsec: calculating radio upper limits}) at points either before or after the time of the high-energy emission, occur during nominal telescope operations, and for which we can calculate radio flux limits $<100$ kJy.

We restrict our radio limits to six hours (21.6 ks) before the high-energy emission from each GRB and 12 hours (43.2 ks) after. We choose this time frame based firstly on the models for FRB-like bursts, which predict most (but not all) pre-GRB radio emission in the seconds before merger and post-GRB emission that could last indefinitely, and secondly on the times at which GRBs are within the FOV of CHIME/FRB. We do not have any SGRBs that are within the FOV of CHIME/FRB in the period of $\sim4$ hours  prior to their high-energy emission.
However, given that pre-SGRB radio emission could commence decades/centuries prior to the merger \citep{Zhang2020}, we arbitrarily extend our range to 6 hours pre-burst. We do not extend our search to the time frame of weeks or years as this is outside the scope of this work.

We list the seven GRBs for which we can constrain the radio emission at the time of the high-energy burst in Table \ref{table: Limits At Time GRB}. Note that while these GRBs are within the FOV of CHIME/FRB, they are not necessarily close to the meridian at the time of their high-energy emission, and thus the limits can be $>$1 kJy. In Table \ref{table: individual GRBs}, we list our most constraining limits for these seven GRBs at times not necessarily equal to that of the high-energy emission, along with the the most constraining limits for the 32 other GRBs.

\begin{deluxetable}{c c c c c}[H]
\tablecaption{CHIME/FRB Upper Limits on Coincident Radio Emission \label{table: Limits At Time GRB}}
\startdata
\\
Name
& Type \tablenotemark{a}
& Flux \tablenotemark{b}
& Fluence Ratio \tablenotemark{c}
& $\eta$ \tablenotemark{d} \\
& & (Jy) & ($10^8$ Jy ms & ($10^{-11}$) \\
& & & erg$^{-1}$ cm$^2$) & 
\\ \hline
GRB 180721A & L & $<$120 & $<$20 & $<$700 \\ 
GRB 181213A & L & $<$1200 & $<$80 & $<$3000 \\ 
GRB 190219A & L & $<$3000 & $<$40 & $<$2000 \\ 
GRB 190613A & L & $<$800 & $<$30 & $<$1000 \\ 
GRB 190613B & L & $<$600 & $<$20 & $<$700 \\ 
GRB 180914A & L & $<$6000 & $<$5 & $<$200 \\ 
GRB 181120A & L & $<$2000 & $<$50 & $<$2000 \\ 
\hline
\enddata
\tablenotetext{a}{Signifies whether the GRB is a SGRB (S) or LGRB (L).}

\tablenotemark{b}{Upper limit on the 400- to 800-MHz radio flux at the 99$\%$ confidence level for a 10-ms radio burst at the time of the high-energy emission (including the dispersion delay).}

\tablenotemark{c}{Upper limit on the 400- to 800-MHz radio-to-high-energy fluence ratio at the 99$\%$ confidence level for a 10-ms radio burst at the time of the high-energy emission (including the dispersion delay).}

\tablenotemark{d}{Upper limit on $\eta$ (unitless radio-to-high-energy fluence ratio assuming a 400-MHz radio bandwidth) at the 99$\%$ confidence level for a 10-ms radio burst at the time of the high-energy emission (including the dispersion delay). The energy range for GRBs detected by \textit{Swift}/BAT is 15 to 150 keV, and that for those detected by \textit{Fermi}/GBM is 10 to 1000 keV.}
\end{deluxetable}

\begin{ThreePartTable}
\begin{TableNotes}
\item\tablenotetext{a}{Signifies whether the GRB is a SGRB (S) or LGRB (L).}
\tablenotemark{b}{Time (in ks) before (negative time) or after the detected high-energy emission for which the radio flux/fluence limits apply. These times include an estimated dispersion delay.}

\tablenotemark{c}{Upper limit on the 400- to 800-MHz radio flux at the 99$\%$ confidence level for a 10-ms radio burst.}

\tablenotemark{d}{Upper limit on the 400- to 800-MHz radio-to-high-energy fluence ratio at the 99$\%$ confidence level for a 10-ms radio burst.}

\tablenotemark{e}{Upper limit on $\eta$ (unitless radio-to-high-energy fluence ratio assuming a 400-MHz radio bandwidth) at the 99$\%$ confidence level for a 10-ms radio burst at the time of the high-energy emission. The energy range for GRBs detected by \textit{Swift}/BAT is 15 to 150 keV, and that for those detected by \textit{Fermi}/GBM is 10 to 1000 keV.}
\end{TableNotes}
\setlength{\tabcolsep}{0.1em}
\begin{longtable}{cccccc}
    \caption{Most Constraining CHIME/FRB Upper Limits on Possible Radio Emission from 39 SGRBs/LGRBs\label{table: individual GRBs}}\\
    \midrule \midrule
  \endfirsthead
\bottomrule
\insertTableNotes
\endlastfoot
Name
& Type\tablenotemark{a}
& Time\tablenotemark{b}
& Flux\tablenotemark{c}
& Fluence Ratio\tablenotemark{d}
& $\eta$\tablenotemark{e}\\
& & (ks) & (Jy) & ($10^8$ Jy ms & ($10^{-11}$) \\
& & & & erg$^{-1}$ cm$^2$) &
\\ 
\midrule 
GRB 181123B & S & $38.016$ & $<$2 & $<$1 & $<$50\\ 
GRB 190326A & S & $39.168$ & $<$2 & $<$4 & $<$140\\ 
GRB 190427A & S & $27.774$ & $<$1.6 & $<$0.9 & $<$40\\ 
GRB 190610A & S & $22.806$ & $<$20 & $<$3 & $<$110\\ 
GRB 190627A & S & $-19.236$ & $<$20 & $<$13 & $<$500\\ 
GRB 180720B & L & $-7.992$ & $<$30 & $<$0.02 & $<$0.6\\ 
GRB 180721A & L & $0.396$ & $<$10 & $<$1 & $<$60\\ 
GRB 180805B & L & $-0.972$ & $<$80000 & $<$5000 & $<$200000\\ 
GRB 180812A & L & $-19.47$ & $<$2 & $<$0.11 & $<$4\\ 
GRB 180818B & L & $16.692$ & $<$2 & $<$0.05 & $<$2\\ 
GRB 180823A & L & $17.232$ & $<$1.5 & $<$0.015 & $<$0.6\\ 
GRB 180904A & L & $20.994$ & $<$2 & $<$0.8 & $<$30\\ 
GRB 180905A & L & $3.924$ & $<$20 & $<$0.3 & $<$11\\ 
GRB 181013A & L & $-21.6$ & $<$400 & $<$130 & $<$5000\\ 
GRB 181023A & L & $-13.11$ & $<$4 & $<$0.6 & $<$20\\ 
GRB 181027A & L & $-18.54$ & $<$3 & $<$0.2 & $<$7\\ 
GRB 181125A & L & $42.726$ & $<$6000 & $<$900 & $<$40000\\ 
GRB 181213A & L & $21.708$ & $<$2 & $<$0.13 & $<$5\\ 
GRB 181224A & L & $-12.318$ & $<$6 & $<$0.8 & $<$30\\ 
GRB 181228A & L & $7.032$ & $<$5 & $<$0.04 & $<$1\\ 
GRB 190103B & L & $-20.892$ & $<$3 & $<$0.02 & $<$0.7\\ 
GRB 190109A & L & $-8.028$ & $<$4 & $<$0.06 & $<$3\\ 
GRB 190109B & L & $-21.078$ & $<$1200 & $<$200 & $<$9000\\ 
GRB 190114A & L & $5.766$ & $<$8 & $<$0.6 & $<$20\\ 
GRB 190202A & L & $30.99$ & $<$3 & $<$0.03 & $<$1\\ 
GRB 190203A & L & $31.764$ & $<$3 & $<$0.014 & $<$0.6\\ 
GRB 190204A & L & $-21.084$ & $<$1100 & $<$7 & $<$300\\ 
GRB 190211A & L & $-18.828$ & $<$2 & $<$0.3 & $<$10\\ 
GRB 190219A & L & $10.83$ & $<$11 & $<$0.1 & $<$6\\ 
GRB 190331A & L & $-17.568$ & $<$5 & $<$1 & $<$40\\ 
GRB 190424A & L & $39.678$ & $<$7 & $<$0.4 & $<$14\\ 
GRB 190515B & L & $25.002$ & $<$3 & $<$0.2 & $<$7\\ 
GRB 190613A & L & $-4.998$ & $<$3 & $<$0.11 & $<$5\\ 
GRB 190613B & L & $0.492$ & $<$30 & $<$0.8 & $<$30\\ 
GRB 190630B & L & $-6.006$ & $<$2 & $<$0.8 & $<$30\\ 
GRB 190701A & L & $19.698$ & $<$4 & $<$0.3 & $<$11\\ 
GRB 180914A & L & $-2.13$ & $<$20 & $<$0.013 & $<$0.5\\ 
GRB 181120A & L & $-3.534$ & $<$10 & $<$0.15 & $<$6\\ 
GRB 190530A & L & $43.188$ & $<$500 & $<$0.12 & $<$5\\ 
GRB 180914B & L & $43.194$ & $<$100 & $<$0.006 & $<$0.2\\  

\end{longtable}   
\end{ThreePartTable}

\begin{figure}[]
    \centering
    \includegraphics[width=0.9\linewidth]{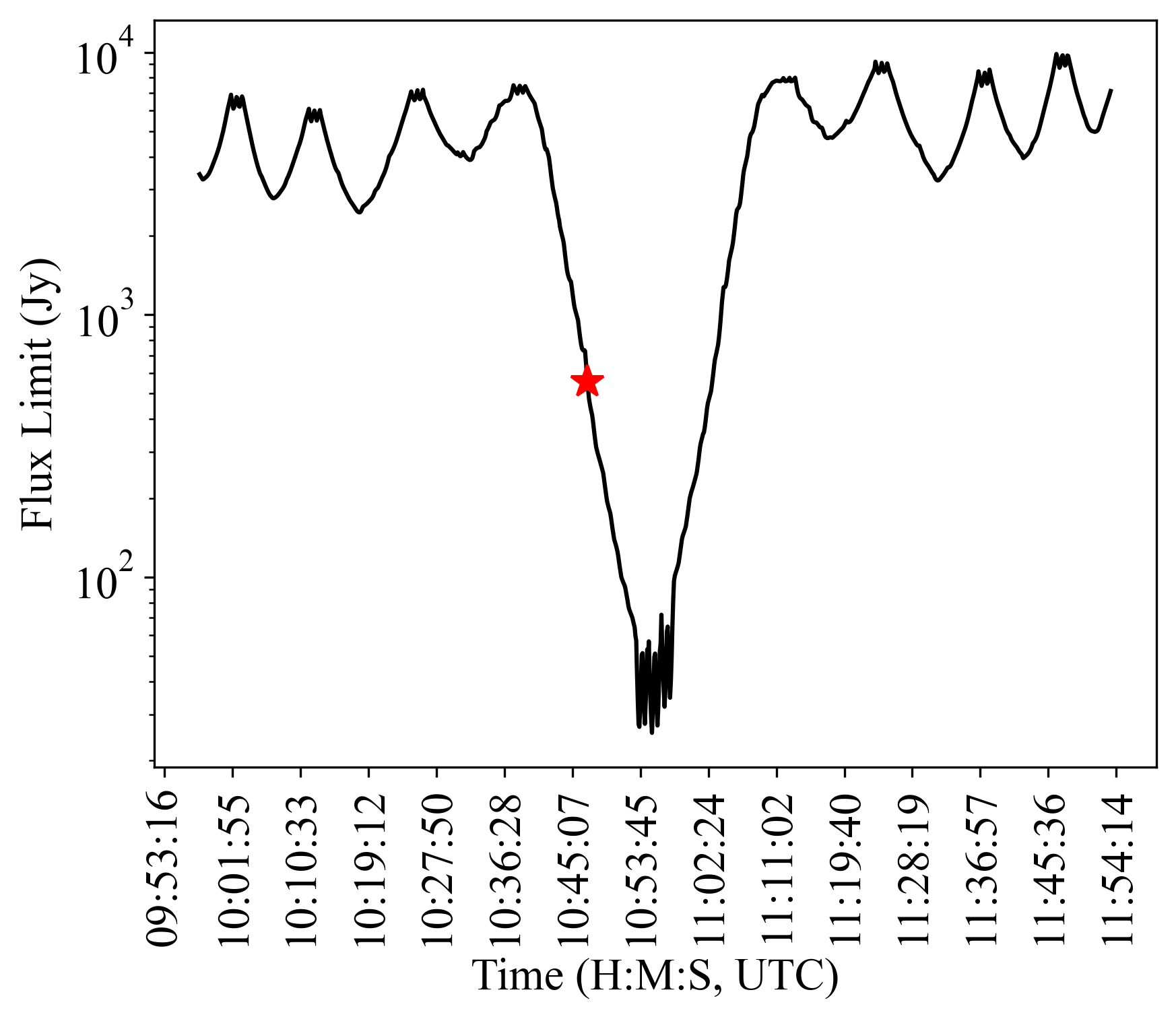}
    \caption{Upper limits on the 400- to 800-MHz radio flux for GRB 190613B assuming a 10-ms radio burst. Radio flux limits (99$\%$ confidence) before and after the high-energy emission (dispersion delay accounted for) are calculated every six seconds. Due to the frequency and continuity of these limits, we show a black solid line rather than showing each individual upper limit. The radio flux limit at the time of the high-energy emission (dispersion delay accounted for) is shown as a red star. All flux density limits are for 2019 June 13, with the hour, minute, and second for which the limit applies shown on the x-axis. The high-energy burst was detected at 10:47:02.05 UT on 2019 June 13 (no dispersion correction) by the \textit{Swift}/BAT telescope. }
    \label{fig:individual GRB}
\end{figure}

In Figure \ref{fig:individual GRB}, we show an example of the 99$\%$ confidence radio flux upper limits for one of our GRBs, GRB 190613B. The limit at the time of the high-energy emission (dispersion delay accounted for) is shown as a red star (and is given in Table \ref{table: Limits At Time GRB}), while limits before and after the GRB (again, dispersion delay accounted for) are depicted using a solid black line. The shape of the limits time series in Figure \ref{fig:individual GRB} is solely due to the transit of the GRB overhead CHIME and our beam shape, with the central minimum value of Figure \ref{fig:individual GRB} given in Table \ref{table: individual GRBs}. 

The evolution of the limits for other GRBs is highly dependent on both the time of the high-energy emission relative to when the source transited directly above CHIME, and the declination of the GRB. Sources closer to the North Celestial Pole are above the horizon at CHIME/FRB for longer than sources with lower declinations e.g., a source at 10$\degree$ in declination is within the primary beam for five minutes while a source at 80$\degree$ is within the primary beam for an hour\footnote{This source would be visible for 30 minutes during an upper transit and 30 minutes during a lower transit for a total of one hour.}. Thus, CHIME/FRB has more exposure to, and thus can calculate limits for longer periods of time, for GRBs at higher rather than lower declinations.

\begin{figure}[]
    \centering
    \includegraphics[width=0.95\linewidth]{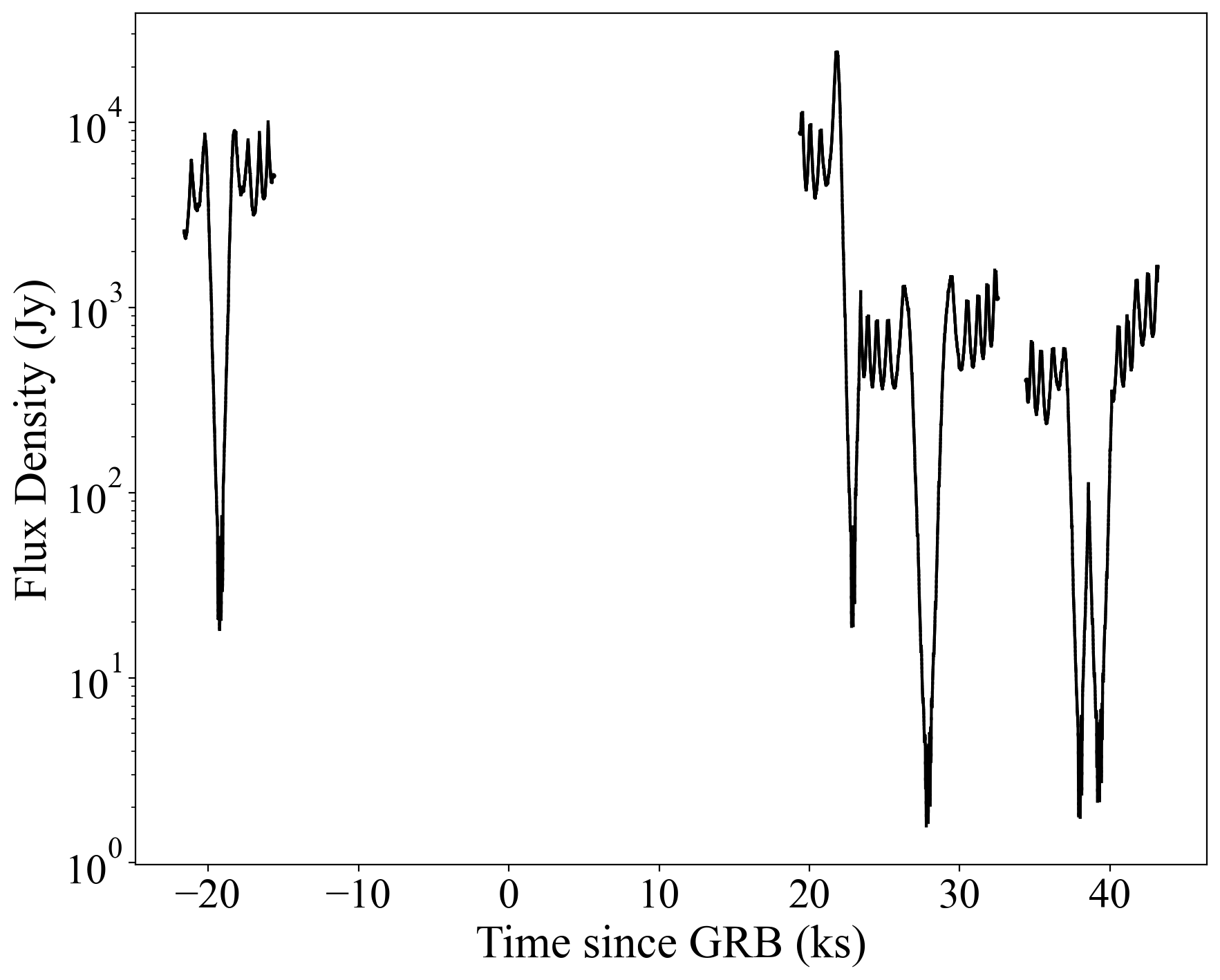}
    \caption{Radio flux limits at the 99$\%$ confidence level for SGRBs within our sample detected by \textit{Swift}/BAT. Upper limits are calculated every six seconds starting six hours (21.6 ks) prior to the GRB until 12 hours (43.2 ks) after the GRB, with the time shifted to account for the necessary DM delay. For each timestamp, we show our most constraining limit from all SGRBs detected by \textit{Swift}/BAT within the FOV of CHIME/FRB at that relative time. }
    \label{fig:SGRBs BAT flux}
\end{figure}

\begin{figure}[]
    \centering
    \includegraphics[width=0.95\linewidth]{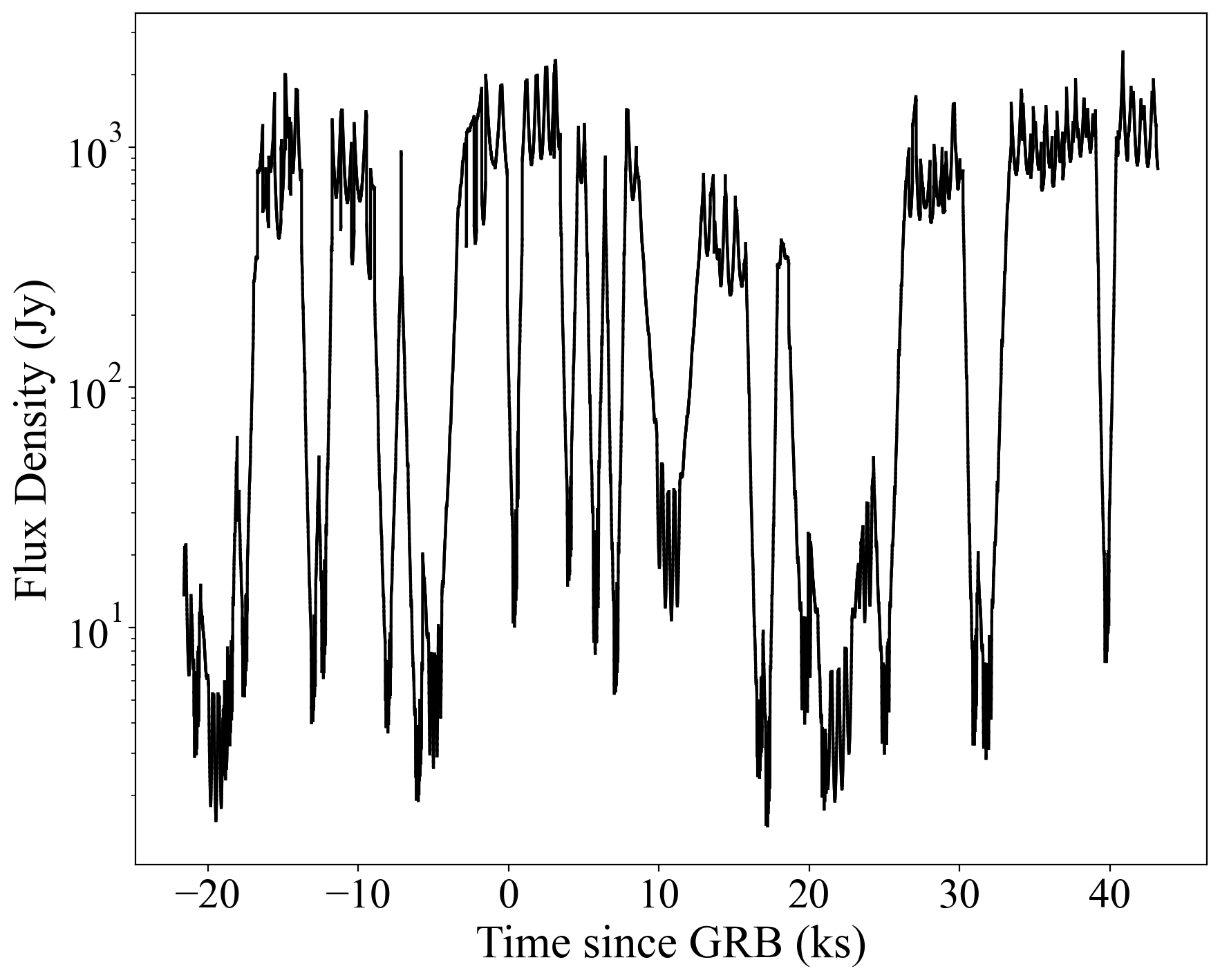}
    \caption{Same as in Figure \ref{fig:SGRBs BAT flux}, except for LGRBs within our sample that are detected by \textit{Swift}/BAT. }
    \label{fig:LGRBs BAT flux}
\end{figure}

Of the 39 GRBs for which we can constrain the radio emission, five are SGRBs detected by \textit{Swift}/BAT, 30 are LGRBs detected by \textit{Swift}/BAT, and four are LGRBs detected by \textit{Fermi}/GBM. Note these classifications are made solely based on the T90s of the bursts. For \textit{Swift}/BAT, we separate the SGRBs from the LGRBs, and show our 3$\sigma$ flux limits (calculated every six seconds) as a function of time relative to the GRB arrival time in Figures \ref{fig:SGRBs BAT flux} and \ref{fig:LGRBs BAT flux}. Rather than showing the flux limit for all GRBs at each relative timestamp, we only show our most constraining FRB-like radio limit at each six second timestamp. Thus, each plot only contains one limit per timestamp with multiple GRBs used to determine the full set of limits shown. Note that a given GRB is often used to determine a handful of radio flux limits for adjacent timestamps, e.g., GRB 181123B is used to place the most constraining SGRB \textit{Swift}/BAT flux limits for $-21.6$ ks to $-17.4$ ks while GRB 190326A is used to place the most constraining flux limits from 20.6 ks to 22.2 ks. Which GRB's limits are the most constraining at a given timestamp, along with how many timestamps the GRB is used for, is dependent on the declination of the GRB and the location of the GRB relative to the meridian at the time of the high-energy emission.

For Figure \ref{fig:SGRBs BAT flux} (and Figure \ref{fig:SGRBs BAT fluence ratios}; discussed below), all five SGRBs in our sample detected by \textit{Swift}/BAT are used to calculate the most constraining flux (and fluence ratio; discussed below) limits for at least one timestamp. However, while all LGRBs in our sample detected by \textit{Swift}/BAT are considered for Figure \ref{fig:LGRBs BAT flux} (and Figure \ref{fig:LGRBs BAT fluence ratios}; discussed below), only a subset of the 30 LGRBs contribute to the most conservative flux (and fluence ratio) limits. These LGRBs are: GRB 190219A, GRB 190701A, GRB 190613B, GRB 181213A, GRB 180818B, GRB 180904A, GRB 180721A, GRB 190211A, GRB 190109A, GRB 190203A, GRB 190613A, GRB 190114A, GRB 190515B, GRB 181228A, GRB 181027A, GRB 181224A, GRB 190331A, GRB 190424A, GRB 190630B, GRB 180823A, GRB 180812A, GRB 190103B, GRB 190202A, GRB 180905A, and GRB 181023A.

Additionally, rather than listing all the limits shown in Figures \ref{fig:SGRBs BAT flux} and \ref{fig:LGRBs BAT flux} in Tables \ref{table: SGRBs Swift} and \ref{table: swift long} (presented in Appendix \ref{sec:Appendix on UL tables}), we give the range of flux limits for a given 1 ks time bin e.g., the minima and maxima of the limits within that time bin. These are the minima and maxima for a given time bin considering all GRBs of that given category (e.g., \textit{Swift}/BAT SGRBs). 

We show our 3$\sigma$ flux limits for LGRBs detected by \textit{Fermi}/GBM\footnote{There are no SGRBs in our sample detected by \textit{Fermi}/GBM.} in Figure \ref{fig:LGRBs Fermi flux} and list the limits in Table \ref{table: LGRBs fermi} in Appendix \ref{sec:Appendix on UL tables}. For each timestamp relative to the GRB, we again only show our most constraining limit from the respective sample of GRBs within a 1 ks time period. Similar to Figure \ref{fig:SGRBs BAT flux}, all LGRBs detected by \textit{Fermi}/GBM in our sample are used to calculate the flux (and fluence ratio) limits for at least one timestamp in Figure \ref{fig:LGRBs Fermi flux} (and Figure \ref{fig:LGRBs Fermi fluence ratios}; discussed below). Note that the large gaps in our upper limits in Figures \ref{fig:SGRBs BAT flux}, \ref{fig:LGRBs Fermi flux}, \ref{fig:SGRBs BAT fluence ratios}, and \ref{fig:LGRBs Fermi fluence ratios} are solely due to a lack of GRBs within the CHIME/FRB FOV at certain timestamps.

In addition to flux limits, we also calculate radio-to-high-energy fluence ratios (both in units of Jy ms erg$^{-1}$ cm$^2$ and as a unitless quantity $\eta$ assuming a 400-MHz bandwidth) using the available high-energy fluences from the online \textit{Swift}/BAT\footnote{https://swift.gsfc.nasa.gov/archive/grb\_table/} and \textit{Fermi}/GBM\footnote{https://heasarc.gsfc.nasa.gov/W3Browse/fermi/fermigbrst.html} catalogs. Similar to our flux limits, we show these ratios for \textit{Swift}/BAT SGRBs, \textit{Swift}/BAT LGRBs, and \textit{Fermi}/GBM LGRBs in Figures \ref{fig:SGRBs BAT fluence ratios}, \ref{fig:LGRBs BAT fluence ratios} and \ref{fig:LGRBs Fermi fluence ratios}, and present them in Tables \ref{table: SGRBs Swift}, \ref{table: swift long}, and \ref{table: LGRBs fermi}, respectively. Here, which GRB's limits are the most constraining at a given timestamp, along with how many timestamps the GRB is used for, is also dependent on the high-energy fluence of the GRB.

\begin{figure}[]
    \centering
    \includegraphics[width=0.95\linewidth]{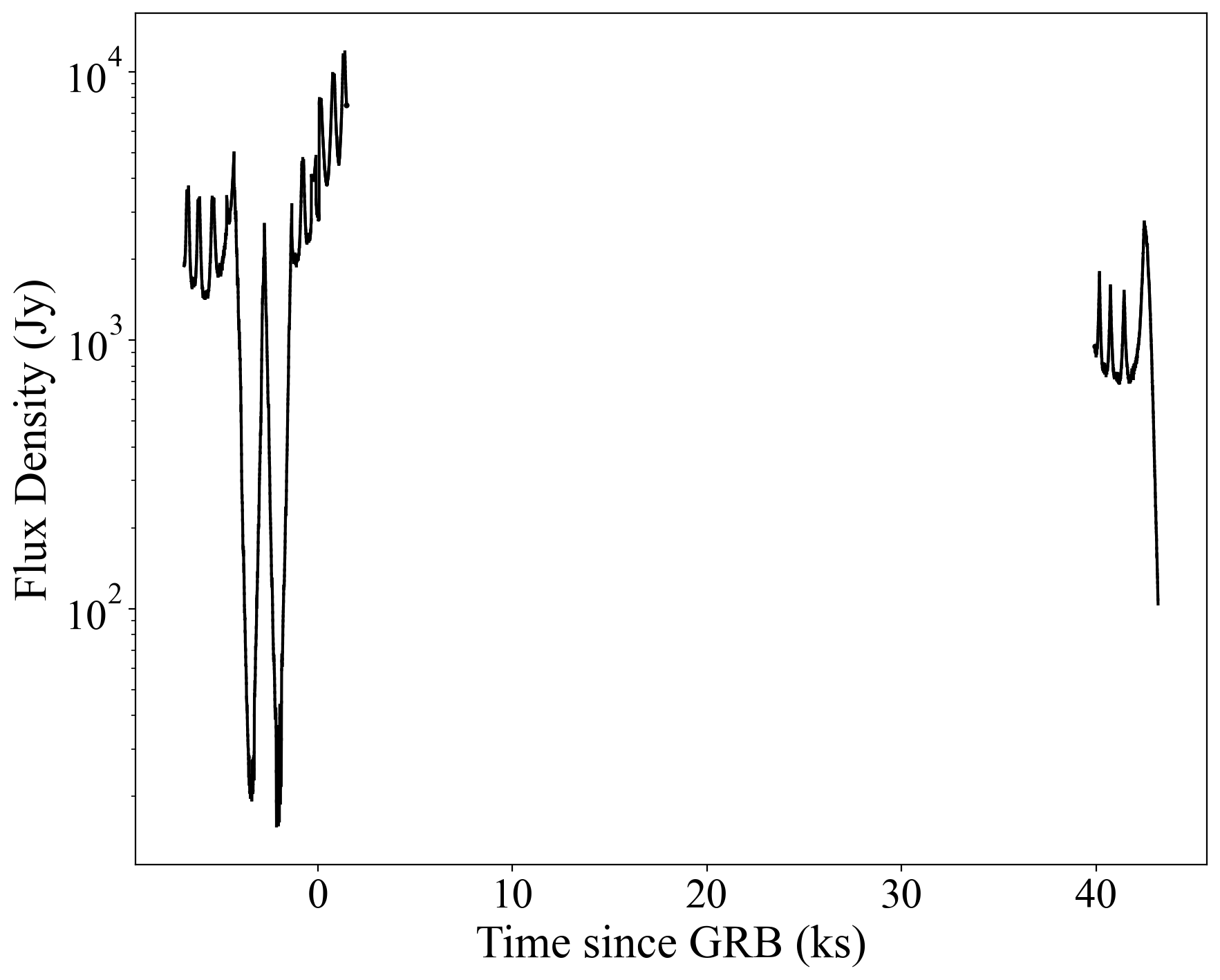}
    \vspace{1em}
    \caption{Same as in Figure \ref{fig:SGRBs BAT flux}, except for LGRBs within our sample that are detected by \textit{Fermi}/GBM.}
    \label{fig:LGRBs Fermi flux}
\end{figure}

\begin{figure}[]
    \centering
    \includegraphics[width=0.95\linewidth]{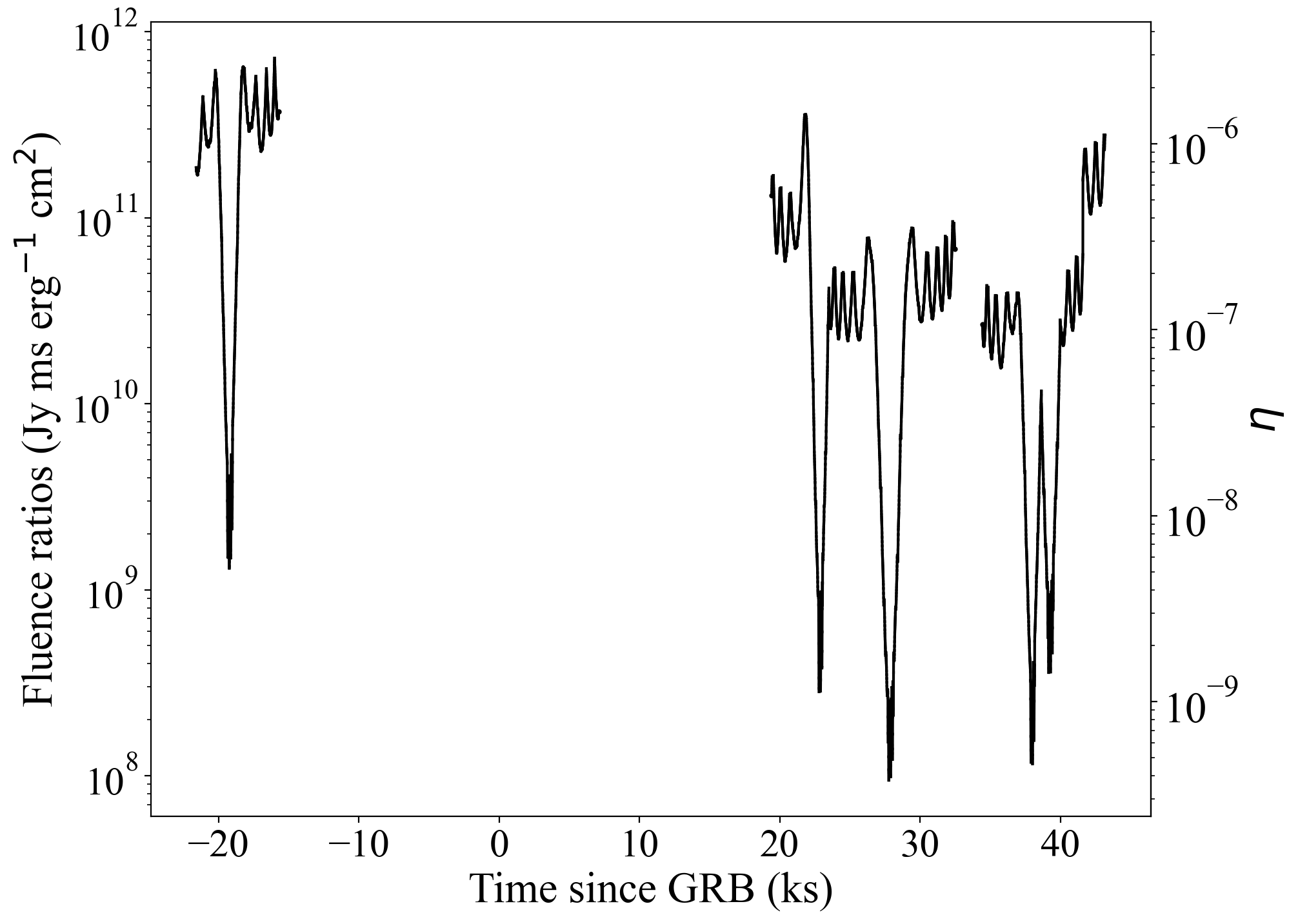}
	\vspace{1em}
    \caption{Same as in Figure \ref{fig:SGRBs BAT flux}, except the radio-to-high-energy fluence ratio limits for SGRBs within our sample that are detected by \textit{Swift}/BAT. We show the fluence ratios in units of Jy ms erg$^{-1}$ cm$^2$ and as unitless quantities ($\eta$) assuming a 400-MHz bandwidth. }
    \label{fig:SGRBs BAT fluence ratios}
\end{figure}

\begin{figure}[]
    \centering
    \includegraphics[width=0.95\linewidth]{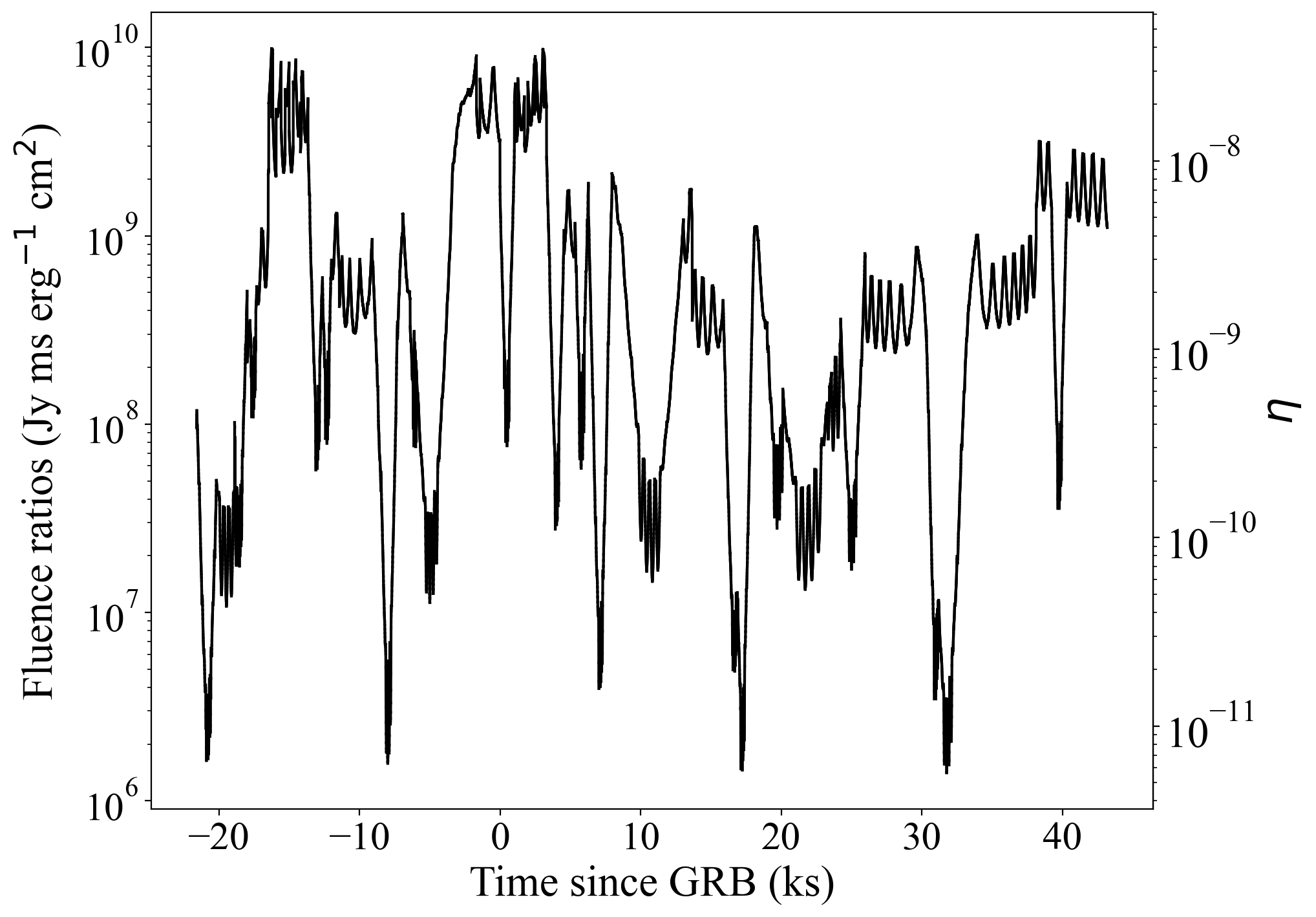}
    \caption{Same as in Figure \ref{fig:SGRBs BAT flux}, except the radio-to-high-energy fluence ratio limits for LGRBs within our sample that are detected by \textit{Swift}/BAT.}
    \label{fig:LGRBs BAT fluence ratios}
\end{figure}

\begin{figure}[]
    \centering
    \includegraphics[width=0.95\linewidth]{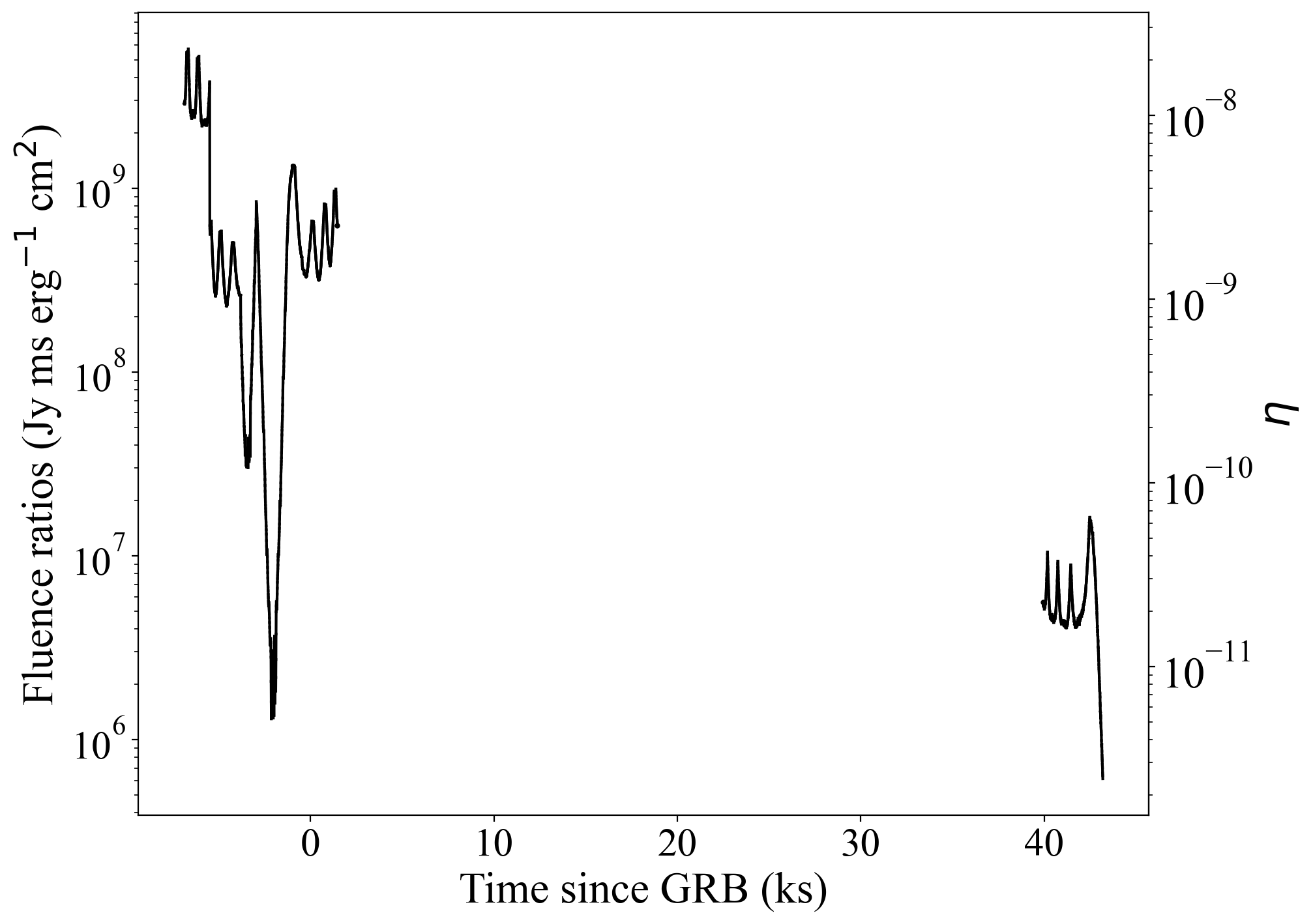}
    \vspace{1em}
    \caption{Same as in Figure \ref{fig:SGRBs BAT flux}, except the radio-to-high-energy fluence ratio limits for LGRBs within our sample that are detected by \textit{Fermi}/GBM. We show the fluence ratios in units of Jy ms erg$^{-1}$ cm$^2$ and as unitless quantities ($\eta$) assuming a 400-MHz bandwidth.}
    \label{fig:LGRBs Fermi fluence ratios}
\end{figure}

\section{Discussion} \label{sec: discussion}

\subsection{SGRBs} \label{sec: discussion SGRBs}
There are a number of models predicting FRB-like radio emission prior to a SGRB, but these models predict radio emission in the $\sim$ms before the NS-NS or NS-BH merger\footnote{We can not place any radio limits at these times as no SGRBs were close enough to the meridian.} \citep[e.g.,][]{Mingarelli2015, Yamasaki2018, Wada2020, Sridhar2021Wind, Carrasco2021}, predict FRB-like radio emission that would be misaligned with the GRB jet \citep[e.g.,][]{2012Piro, Wang2016,Sridhar2021Wind}, or predict radio emission that would be too weak to be visible by CHIME/FRB \citep[e.g.,][]{2012Piro, Wang2016}. The model that predicts FRB-like radio emission that might be detectable by CHIME/FRB is that by \citet{Zhang2020} in which FRB-like radio emission is generated prior to the merger of two NS due to interactions between their magnetospheres. This emission would start decades to centuries prior to the merger, with the bursting activity increasing as the two NSs inspiral. While there is not any quantitative prediction for the radio flux of this model, it is possible that our limits could constrain it, with our best pre-SGRB limit of $<20$ Jy at $\sim$5 hours (18.6 ks) prior to the high-energy emission (see Table \ref{table: SGRBs Swift}).

\begin{figure}[]
    \centering
    \includegraphics[width=0.95\linewidth]{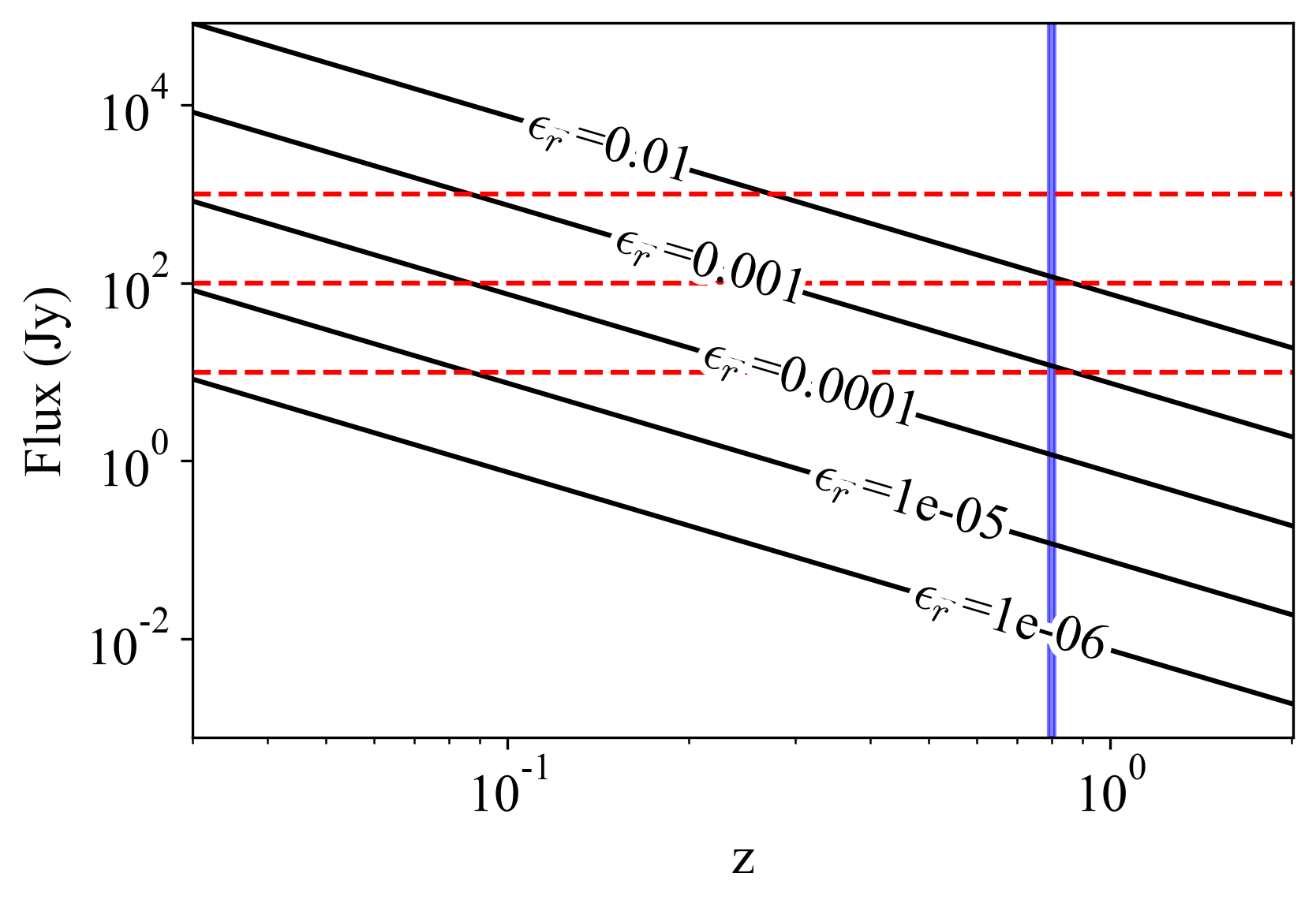}
    \vspace{1em}
    \caption{Radio flux densities predicted by \citet{Totani2013} for FRB-like emission after a NS-NS merger. Flux densities at 600-MHz are shown as a function of redshift, with different diagonal lines corresponding to different radio efficiencies in Eq.~\eqref{eq: totani}. The three red, dashed, horizontal lines show flux limits for an FRB-like radio burst of 10 Jy, 100 Jy, and 1 kJy. The blue line is placed at the median SGRB redshift, $\textrm{z}=0.82$, from the online GRB catalog GRBWeb \citep[][]{GRBweb}.}
    \label{fig:Totani simulation}
\end{figure}

After the SGRB, if the merger results in a NS, then both \citet{PshirkovPostnov2010} and \citet{Totani2013} predict that FRB-like emission is possible from magnetic braking in a manner similar to that of pulsar emission. As derived by \citet{Totani2013} and then expanded on by \citet{RowlinsonAnderson2019}, the predicted flux density from this pulsar-like emission is:

\begin{equation}
    F_{\nu} = 8 \times 10^7 \nu_{\textrm{obs}}^{-1} \epsilon_r D^{-2} B_{15}^2 R_6^6 P_{-3}^{-4} \textrm{Jy}
    \label{eq: totani}
\end{equation}
\noindent
where $\nu_{\textrm{obs}}$ is the observing frequency in MHz, $\epsilon_r$ the efficiency of converting energy into radio emission, $D$ the distance in Gpc, $B_{15}$ the magnetic field strength in units of $10^{15}$ G, $R_6$ the radius of the NS in units of $10^6$ cm, and $ P_{-3}$ the NS spin period in units of $10^{-3}$ s. While all of these factors except for observing frequency are unknown for our sample of GRBs\footnote{One of the SGRBs in our sample, GRB 190627A, has a redshift of 1.94. However, this SGRB was only within the FOV of CHIME/FRB prior to the high-energy emission, and thus cannot be used to constrain post-SGRB radio emission models.}, we assume a millisecond magnetar and hence set $B_{15} = 1$, $R_6 = 1$, and $ P_{-3} = 1$. We show the predicted flux density for a range of redshifts and $\epsilon_r$ in Figure \ref{fig:Totani simulation}. While our chosen parameters describe a millisecond magnetar, a slower magnetar could generate the same level of radio emission if the magnetic field strength were appropriately scaled up. Note that our limits assume a burst width of 10-ms, which is unrealistically long for a millisecond magnetar. However, as the flux scales by the square root of the width, this will only result in a factor of $\sim$3 difference in our flux/fluence limits.

Assuming the surrounding environment is transparent to MHz radio emission at $\sim$8 to 12 hours (28.8 to 43.2 ks) post-SGRB, we set limits on the radio efficiency, $\epsilon_r$, of Eq. ~\eqref{eq: totani} using some of our most constraining post-SGRB limits. We list the time periods for which we can constrain the radio emission to $<$ 10 Jy, $<$ 100 Jy, and $<$ 1 kJy in Table \ref{table: epsilon r limits}. Additionally, using our above assumptions, we calculate constraints on $\epsilon_r$ for this model assuming two different redshifts for our SGRBs, $z<1$ and $z<0.1$. Note that of the 74 SGRBs in the online GRB catalog GRBWeb \citep[][]{GRBweb}, 46 have redshifts $<1$ and six have redshifts $<0.1$. While we are able to constrain $\epsilon_r$ over $\sim$hour-long timescales, it is possible that a radio burst from a GRB could either be a) fainter than our limits or b) occur outside the time frame of our limits. Thus $\epsilon_r$ is only constrained by our work for very specific timestamps.

\begin{deluxetable*}{l l l l | l l | l l l}
\setlength{\tabcolsep}{4pt}
\tablecaption{Duration of SGRB Flux Limits and Model Constraints} \label{table: epsilon r limits}
\startdata
\\
Flux Limit & 
Start time\tablenotemark{a} & 
End time\tablenotemark{b} & 
Total time \tablenotemark{c} & 
$\epsilon_{r}$ ($z < 1$) \tablenotemark{d}& 
$\epsilon_{r}$ ($z < 0.1$) \tablenotemark{e}& 
$\epsilon_{r}$ ($z < 1$) \tablenotemark{f}& 
$\epsilon_{r}$ ($z < 0.1$) \tablenotemark{g}& 
$\alpha$\tablenotemark{h}\\ 
(Jy) & (ks) &  (ks) & (min.) & (Totani model) & (Totani model) & (Zhang model)& (Zhang model) &\\\hline
$<10$ & \begin{tabular}[c]{@{}l@{}}27.498 \\ 37.746  \\ 38.976 \end{tabular} & \begin{tabular}[c]{@{}l@{}}28.200 \\ 38.208 \\ 39.558  \end{tabular} & 29.2 & $<0.0013$ & $< 1.3 \times 10^{-5}$ & \begin{tabular}[c]{@{}l@{}} $<9.5 \times 10^{-10}$\\ $<2.8 \times 10^{-4}$ \\ $<110$ \end{tabular} & \begin{tabular}[c]{@{}l@{}} $<5.2 \times 10^{-12}$ \\ $<1.6 \times 10^{-6}$ \\ $<0.6$\end{tabular} &\begin{tabular}[c]{@{}l@{}} $-2$ \\ $-3$ \\ $-4$\end{tabular}\\ \hline
$<100$ & \begin{tabular}[c]{@{}l@{}}$-19.446$\\ 22.662 \\ 27.096 \\ 37.386 \end{tabular} & \begin{tabular}[c]{@{}l@{}}$-19.002$\\ 23.094 \\ 28.590 \\  39.936 \end{tabular} & 82.1  & $<0.013$ & $<1.6 \times 10^{-4}$ & \begin{tabular}[c]{@{}l@{}} $<9.5 \times 10^{-9}$\\ $<2.8 \times 10^{-3}$ \\ $<1100$ \end{tabular} & \begin{tabular}[c]{@{}l@{}} $<5.2 \times 10^{-11}$ \\ $<1.6 \times 10^{-5}$ \\ $<6$\end{tabular}&\begin{tabular}[c]{@{}l@{}} $-2$ \\ $-3$ \\ $-4$\end{tabular}\\ \hline
$<1000$ &\begin{tabular}[c]{@{}l@{}}$-19.770$\\ 22.350 \\ 26.544 \\ 29.628 \\ 34.446 \\ 41.916 \\42.672 \end{tabular} & \begin{tabular}[c]{@{}l@{}}$-18.660$\\ 26.154 \\29.148 \\ 32.224 \\ 41.622 \\ 42.384 \\43.002 \end{tabular} & 294.9 & $<0.13$ & $<1.6 \times 10^{-3}$& \begin{tabular}[c]{@{}l@{}} $<9.5 \times 10^{-8}$\\ $<2.8 \times 10^{-2}$ \\ $<11000$ \end{tabular} & \begin{tabular}[c]{@{}l@{}} $<5.2 \times 10^{-10}$ \\ $<1.6 \times 10^{-4}$ \\ $<60$\end{tabular}&\begin{tabular}[c]{@{}l@{}} $-2$ \\ $-3$ \\ $-4$\end{tabular}\\ \hline
\enddata
\tablenotetext{a}{Start time for a specific time segment for which we can constrain the flux to be less than that listed in the first column of the table e.g., 10 Jy, 100 Jy, or 1 kJy. A positive time indicates a time post-high-energy emission, while a negative time indicates a time pre-high-energy emission.}
\tablenotetext{b}{ End time for this specific time segment for which we can constrain the flux to be less than that listed in the first column of the table e.g., 10 Jy, 100 Jy, or 1 kJy. A positive time indicates a time post-high-energy emission, while a negative time indicates a time pre-high-energy emission.}
\tablenotemark{c} {Total time over the entire time period of six hours prior and 12 hours after the high-energy emission for which we can constrain the flux to be less than the limit given in the first column.}
\tablenotetext{d} {Constraints on $\epsilon_r$ in the model by \citet{Totani2013} assuming $z<1$ for the SGRBs used to determine the limits. For this model, we also assume that the GRB would be have the following properties: $B_{15} = 1$, $R_6 = 1$, and $ P_{-3} = 1$.}
\tablenotetext{e} {Same as in \textit{c} except for $z < 0.1$.}
\tablenotetext{f} {Constraints on $\epsilon_r$ in the model by \citet{zhang2014} assuming the redshifts of the SGRBs used to determine the limits are $z<1$. We list the $\alpha$ used for the calculation in the final column. For this model, we also assume the following properties: $E_B = 1.7 \times 10^{47}$ erg, $\tau = 1$ ms, and $\nu_p = 1$ kHz, and $\nu_{\textrm{obs}}=600$ MHz for the other model parameters.}
\tablenotetext{g} {Same as in \textit{e} except for $z < 0.1$.}
\tablenotetext{h} {The value for $\alpha$ in the model by \citet{zhang2014} which is used to calculate the $\epsilon_r$ values in the previous two columns.}
\end{deluxetable*}

It is also possible that at our times of interest the surrounding environment was not yet transparent to radio emission. This has been suggested by \citet{Yamasaki2018}, who find that the ejecta surrounding a remnant magnetar from a NS-NS merger would not be transparent to GHz radio emission for $\sim$years post-merger (although there is significant uncertainty on this estimation). Additionally, while it is suggested that the FRB and GRB emission would be emitted along the same axis in this model \citep[]{RowlinsonAnderson2019}, it is possible that the two are emitted along different axes, and hence the radio and high-energy emission would not both be observable from the same source. Lastly, it is possible that the duration of the radio burst could be longer than the CHIME/FRB parameter search space, and hence would not be detectable by CHIME/FRB. While CHIME/FRB searches for bursts out to $\sim$100 ms \citep[e.g., see][]{abb+18}, the sensitivity is greatly reduced for bursts of this duration \citep[see Figure 7 of][]{2022Merryfield}, and hence it is more reasonable to assume CHIME/FRB is senstive to bursts up to $\sim$10-30 ms.

In addition to pulsar-like radio emission from a NS-NS merger, radio emission might also be produced if the resulting NS is unstable and collapses to form a BH \citep{FalckeRezzolla2014, zhang2014}. This might occur hundreds to thousands of seconds post merger, which is within the time frame of our SGRB limits. 

The flux density of the radio emission predicted by \citet{zhang2014} and further expanded on by \citet{RowlinsonAnderson2019} can be described by:

\begin{equation}
    F_{\nu} = \frac{-10^{-23} \epsilon_r E_B}{4 \pi D^2 \tau} (\alpha +1) \nu_p^{-(\alpha+1)} \frac{\nu_{\textrm{obs}}^{\alpha}}{(1+z)} \textrm{Jy}
    \label{eq: zhang}
\end{equation}
\noindent
where $\epsilon_r$ is the efficiency of converting the magnetic energy into radio emission, $E_B$ is the predicted magnetic energy released during the magnetic reconnection event responsible for the FRB, $D$ is the distance in Gpc, $\tau$ is the time over which the magnetic energy is released, $\nu_p$ is the plasma frequency in MHz, $\alpha$ is the spectral index of the emission, $\nu_{\textrm{obs}}$ is the observing frequency in MHz, and $z$ is the redshift of the GRB. Note that for the flux to be positive, $\alpha < -1$. 

Again, almost all of these parameters are unknown for our sample of GRBs. Thus, we assume the parameters presented by \citet{zhang2014} and \citet{RowlinsonAnderson2019} of $E_B = 1.7 \times 10^{47}$ erg, $\tau = 1$ ms, and $\nu_p = 1$ kHz. We assume $\nu_{\textrm{obs}}=600$ MHz for CHIME/FRB. The three remaining unknowns are $\alpha$, $z$, and $\epsilon_r$. 

\begin{figure}[]
    \centering
    \includegraphics[width=0.95\linewidth]{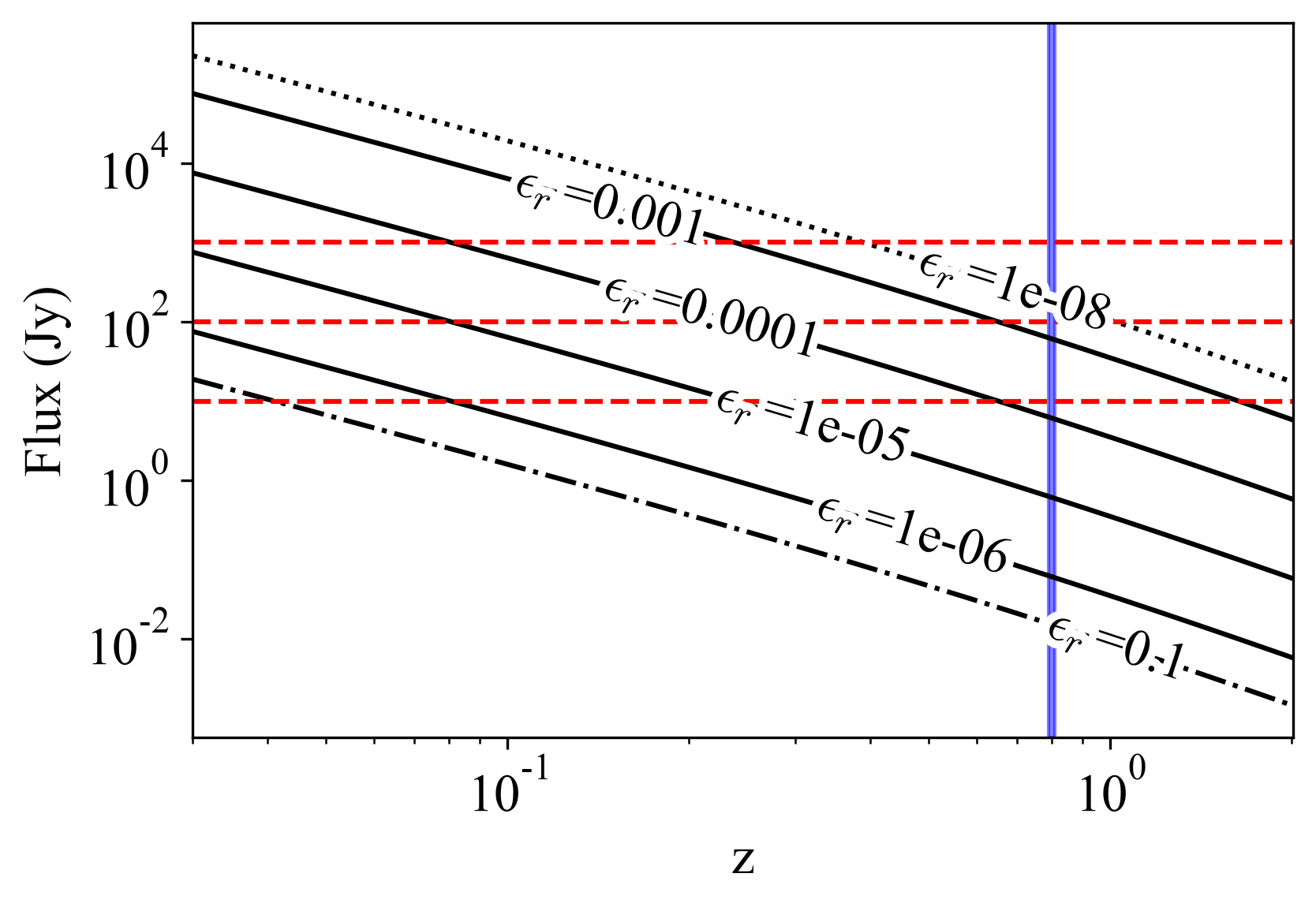}
    \vspace{1em}
    \caption{Radio flux densities predicted by \citet{zhang2014} for FRB-like emission during the collapse of a NS to a BH. Flux densities are shown as a function of redshift, with different diagonal lines corresponding to different radio efficiencies in Eq.~\eqref{eq: zhang}. The solid black lines show the flux densities for $\alpha=-3$, while the dotted line is for $\alpha=-2$, and the dashed dotted line is for $\alpha=-4$. The three red dashed, horizontal lines show flux limits for an FRB-like radio burst of 10 Jy, 100 Jy, and 1 kJy. The blue line is placed at the median SGRB redshift, $\textrm{z}=0.82$ from the online GRB catalog GRBWeb \citep{GRBweb}.}
    \label{fig:zhang simulation}
\end{figure}

We list in Table \ref{table: epsilon r limits} the limits on $\epsilon_r$ for our 10 Jy, 100 Jy, and 1 kJy limits. We originally calculate all the constraints using $\alpha =-3$ as used by \citet{RowlinsonAnderson2019}, and show the predicted flux as a function of redshift for $\alpha=-3$ as the solid lines in Figure \ref{fig:zhang simulation}. However, as there is no strong evidence for $\alpha =-3$, we also calculate the limits for $\alpha=-2$ and $\alpha=-4$. Assuming $\alpha=-2$, our limits become even more constraining (e.g., see the dotted line in Figure \ref{fig:zhang simulation}). Conversely, if $\alpha=-4$, our limits are significantly less constraining (e.g., see the dashed dotted line in Figure \ref{fig:zhang simulation}). 

If the SGRB is produced by a BH-NS merger rather than a NS-NS merger, then \citet{Mingarelli2015} predict emission similar to that of \citet{zhang2014} from the transfer of the NS's magnetic field to the BH during the merger. While this emission might occur as soon as $\sim$milliseconds after the merger, it is possible that the BH might be able to maintain the magnetic fields of the NS for a significant period of time, and thus the FRB-like burst might be produced a short while after the merger\footnote{The maximum length of time for which the BH could maintain the magnetic fields of the NS has not yet been quantified, but it is at least 0.5 ms \citep{Mingarelli2015}.}. However, with a predicted luminosity of $10^{41}$ erg s$^{-1}$, the radio flux densities from this model would be significantly below the limits for our current sample unless the SGRB redshifts were $<0.08$. Of the 74 SGRBs with redshifts in the online GRB catalog GRBWeb \citep[][]{GRBweb}, only three have redshifts $<0.08$, and thus this is a very small fraction of the SGRB population. 

\subsection{LGRBs}
There are fewer theoretical predictions for radio emission coincident with LGRBs. The recent detection of an LGRB associated with a kilonova (and hence a compact object merger) by \citet{2022LGRBKilonova} and \citet{2022YangLGRBMerger} suggests that some of the emission mechanisms mentioned in Section \ref{sec: discussion SGRBs} might be relevant for LGRBs. However, we leave this to future work, as hopefully more LGRBs associated with compact object mergers will be detected and a distinction can be made between the LGRBs that are produced in this manner versus in the more classical manner of a core-collapse supernova. 

For LGRBs not associated with compact object mergers, it is likely that the surrounding ejecta of the young, magnetized NS produced by the supernova would not be transparent to radio emission for $\sim$ years to decades post-LGRB, and thus the radio emission would not have been detectable during our observing periods \citep{mbm17}. 

However, given our sample of LGRBs for which we can place radio emission limits is significantly larger than our sample of SGRBs\footnote{There are 34 LGRBs for which we can constrain the radio flux and only 5 SGRBs for which we can do this.}, we still calculate radio flux limits for our set of LGRBs. These limits span the full period of 6 hours (21.6 ks) prior to and 12 hours (43.2 ks) after the time of the GRB, with limits as small as 2 Jy in the $\sim$5 ks before the high-energy emission (see Table \ref{table: swift long} and \ref{table: LGRBs fermi}). 

\section{Summary and Future Work}
\label{sec: summary}
In this work, we have searched for temporally and spatially coincident FRBs and GRBs using the first CHIME/FRB catalog along with all well-localized GRBs detected between 2018 July 17 and 2019 July 8. We do not find any temporal and spatial coincidences within the combined 3$\sigma$ uncertainty regions of the FRBs and GRBs, and a time range of one week. There are two GRBs that are solely spatially coincident with two FRBs in our sample within a time frame longer than one week, but the chance of such a coincidence is high, and thus we do not find these coincidences to be significant.

In addition to checking for coincident GRBs and FRBs, we also determine upper limits on possible radio emission for 39 GRBs either before, at the time of, or after the time of high-energy emission. We can constrain the radio emission for SGRBs to be $<10$ Jy for a total of 29.2 minutes, $<100$ Jy for a total of 82.1 minutes, and $<1$ kJy for a total of 294.9 minutes.

Unfortunately, there are no SGRBs within the FOV of CHIME/FRB at the time of the high-energy emission or within a few ks of it. Assuming our SGRB sample is associated with NS-NS mergers, our 10 Jy limits can constrain the radio efficiency of the emission produced by a magnetar in the model of \citet{Totani2013} to be $<0.0013$ if our GRB is at $z < 1$ and $<1.3\times10^{-5}$ if the GRB is at $z < 0.1$. Additionally, if the remnant magnetar is unstable and collapses to form a BH in a manner similar to that outlined by \citet{zhang2014}, then using our 10 Jy limits we can additionally constrain the radio efficiency of the emission to be $<2.8 \times 10^{-4}$ for $\alpha = -3$ if the GRB is at $z < 1$, or $<1.6 \times 10^{-6}$ for $\alpha = -3$ if the GRB is at $z < 0.1$. Our larger radio limits are less constraining, but still interesting as they are relevant over longer periods of time. However, we note that it is still possible for radio bursts to occur just outside our time windows with unconstrained $\epsilon_r$.

For LGRBs, we can constrain FRB-like radio emission to $<10$ Jy for a total of 91.3 minutes pre-LGRB and a total of 98.7 minutes post-LGRB, but there are no strong theoretical predictions for coincident FRB-like radio emission for LGRBs. Within 0.1 ks of the high-energy emission, our best constraint is $<$25 Jy for a 10-ms radio burst (with this limit valid for $\sim$ 4.4 minutes). However, given the limited number of theories predicting radio emission at the time of an LGRB, we do not use our LGRB limits to constrain any radio-emission mechanisms. Future work which could explore using LGRB limits to possibly constrain radio emission from compact object mergers.

Given the large number of FRBs that have been and will be detected by the CHIME/FRB collaboration, future work will continue to search for coincident FRBs and GRBs using the most-recent GRB catalogs along with all the verified FRBs detected by CHIME/FRB. Additionally, it is possible that more recent SGRBS (e.g., outside the time frame considered here) have been within the main lobe of CHIME/FRB at the time of their high-energy emission. Whether or not future radio emission will be detected from SGRBs by CHIME/FRB depends both on the chance of an SGRB being within the mainlobe of CHIME/FRB at the time of the high-energy detection (which did not occur for the sample considered here) and the SGRB emitting radio emission that is temporally coincident and beamed in the same direction as the high-energy emission. It is possible that a detection of radio emission at a time different than that of the high-energy emission is also possible, but it will be harder to associate with a given GRB due to high chance coincidences. Nonetheless, it remains of interest to continue searching for FRB-GRB coincidences.

\section*{acknowledgments}
We acknowledge the use of public data from the \textit{Swift} data archive and the \textit{Fermi} data archive. We acknowledge that CHIME is located on the traditional, ancestral, and unceded territory of the Syilx/Okanagan people. We are grateful to the staff of the Dominion Radio Astrophysical Observatory, which is operated by the National Research Council of Canada.  CHIME is funded by a grant from the Canada Foundation for Innovation (CFI) 2012 Leading Edge Fund (Project 31170) and by contributions from the provinces of British Columbia, Qu\'{e}bec and Ontario. The CHIME/FRB Project is funded by a grant from the CFI 2015 Innovation Fund (Project 33213) and by contributions from the provinces of British Columbia and Qu\'{e}bec, and by the Dunlap Institute for Astronomy and Astrophysics at the University of Toronto. Additional support was provided by the Canadian Institute for Advanced Research (CIFAR), McGill University and the McGill Space Institute thanks to the Trottier Family Foundation, and the University of British Columbia.

\allacks

\bibliography{refs, frbrefs}
\bibliographystyle{aasjournal}

\appendix

\section{Upper Limits as a Function of Time for SGRBs and LGRBs}
\label{sec:Appendix on UL tables}

Below, in Tables \ref{table: SGRBs Swift}, \ref{table: swift long}, and \ref{table: LGRBs fermi}, we present our most constraining radio flux limits and fluence ratios for SGRBs detected by \textit{Swift}/BAT, LGRBs detected by \textit{Swift}/BAT, and LGRBs detected by \textit{Fermi}/GBM. For each kilosecond timestamp, we give the range of our most constraining upper limits calculated considering the entire sample of GRBs within that category (e.g., for Table \ref{table: SGRBs Swift}, all SGRBs detected by \textit{Swift}/BAT) are used to determine the limits.

\begin{ThreePartTable}
\begin{TableNotes}
\item\tablenotemark{a}{End time of a given ks bin for which the radio flux/fluence limit ranges apply. Times are relative to the detected high-energy emission, with negative times indicating emission prior to the high-energy emission. Start time for the range of fluxes/fluences is 1 ks prior to the end time.}

\tablenotemark{b}{Range of upper limits on the possible radio flux at the 99$\%$ confidence level for a 10-ms radio burst for the given ks time bin. The flux range is calculated using all \textit{SGRBs} detected by \textit{Swift}/BAT within our sample.}

\tablenotemark{c}{Range of upper limits on the possible radio-to-high-energy fluence ratio at the 99$\%$ confidence level for a 10-ms radio burst for the given ks time bin. The fluence ratio range is calculated using all \textit{SGRBs} detected by \textit{Swift}/BAT within our sample.}

\tablenotemark{d}{Range of upper limits on $\eta$ (unitless radio-to-high-energy fluence ratio assuming a 400-MHz radio emission bandwidth) at the 99$\%$ confidence level for a 10-ms radio burst for the given ks time bin. The fluence ratio range is calculated using all \textit{SGRBs} detected by \textit{Swift}/BAT within our sample. The \textit{Swift}/BAT high-energy band is 15 to 150 keV, while the CHIME/FRB radio band is 400- to 800-MHz.}
\end{TableNotes}
\begin{longtable}{cccc}
    \caption{CHIME/FRB Upper Limits on Radio Emission from SGRBs detected by  \textit{Swift}/BAT}
    \label{table: SGRBs Swift}\\
    \midrule \midrule
  \endfirsthead

\bottomrule
\insertTableNotes
\endlastfoot

Time \tablenotemark{a}
& Flux \tablenotemark{b}
& Fluence Ratio \tablenotemark{c}
& $\eta$\tablenotemark{d} \\
(ks) & (Jy) & ($10^8$ Jy ms & ($10^{-11}$)\\
 & & erg$^{-1}$ cm$^2)$ &
\\
\midrule
$-20.6$ & $<$2000-6000 & $<$2000-4000 & $<$70000-200000\\ 
$-19.6$ & $<$300-9000 & $<$200-6000 & $<$7000-200000\\ 
$-18.6$ & $<$20-2000 & $<$13-1300 & $<$500-50000\\ 
$-17.6$ & $<$2000-9000 & $<$1400-6000 & $<$60000-300000\\ 
$-16.6$ & $<$3000-8000 & $<$2000-6000 & $<$90000-200000\\ 
$-15.6$ & $<$4000-10000 & $<$3000-7000 & $<$110000-300000\\ 
$20.4$ & $<$4000-11000 & $<$600-2000 & $<$20000-70000\\ 
$21.4$ & $<$4000-9000 & $<$600-1400 & $<$20000-50000\\ 
$22.4$ & $<$700-20000 & $<$110-4000 & $<$4000-140000\\ 
$23.4$ & $<$20-1100 & $<$3-200 & $<$110-6000\\ 
$24.4$ & $<$400-1200 & $<$200-500 & $<$7000-20000\\ 
$25.4$ & $<$400-900 & $<$200-500 & $<$9000-20000\\ 
$26.4$ & $<$400-1300 & $<$200-800 & $<$9000-30000\\ 
$27.4$ & $<$16-1200 & $<$10-700 & $<$400-30000\\ 
$28.4$ & $<$1-30 & $<$0.9-15 & $<$40-600\\ 
$29.4$ & $<$20-1500 & $<$14-900 & $<$600-30000\\ 
$30.4$ & $<$500-1500 & $<$300-900 & $<$11000-40000\\ 
$31.4$ & $<$500-1200 & $<$300-700 & $<$11000-30000\\ 
$32.4$ & $<$500-1600 & $<$300-1000 & $<$13000-40000\\ 
$33.4$ & $<$1100-1600 & $<$700-900 & $<$30000-40000\\ 
$35.4$ & $<$300-700 & $<$200-400 & $<$7000-20000\\ 
$36.4$ & $<$200-600 & $<$160-400 & $<$6000-16000\\ 
$37.4$ & $<$90-600 & $<$60-400 & $<$2000-16000\\ 
$38.4$ & $<$2-90 & $<$1-60 & $<$50-2000\\ 
$39.4$ & $<$2-110 & $<$4-120 & $<$140-5000\\ 
$40.4$ & $<$3-500 & $<$5-300 & $<$200-12000\\ 
$41.4$ & $<$400-900 & $<$200-600 & $<$10000-20000\\ 
$42.4$ & $<$500-1400 & $<$300-2000 & $<$12000-90000\\ 
\end{longtable}   
\end{ThreePartTable}

\FloatBarrier

\begin{ThreePartTable}
\begin{TableNotes}
\item\tablenotemark{a}{End time of a given ks bin for which the radio flux/fluence limit ranges apply. Times are relative to the detected high-energy emission, with negative times indicating emission prior to the high-energy emission. Start time for the range of fluxes/fluences is 1 ks prior to the end time.}

\tablenotemark{b}{Range of upper limits on the possible radio flux at the 99$\%$ confidence level for a 10-ms radio burst for the given ks time bin. The flux range is calculated using all \textit{LGRBs} detected by \textit{Swift}/BAT within our sample.}

\tablenotemark{c}{Range of upper limits on the possible radio-to-high-energy fluence ratio at the 99$\%$ confidence level for a 10-ms radio burst for the given ks time bin. The fluence ratio range is calculated using all \textit{LGRBs} detected by \textit{Swift}/BAT within our sample.}

\tablenotemark{d}{Range of upper limits on $\eta$ (unitless radio-to-high-energy fluence ratio assuming a 400-MHz radio emission bandwidth) at the 99$\%$ confidence level for a 10-ms radio burst for the given ks time bin. The fluence ratio range is calculated using all \textit{LGRBs} detected by \textit{Swift}/BAT within our sample. The \textit{Swift}/BAT high-energy band is 15 to 150 keV, while the CHIME/FRB radio band is 400- to 800-MHz.}
\end{TableNotes}
\begin{longtable}{cccc}
    \caption{CHIME/FRB Upper Limits on Radio Emission from LGRBs detected by \textit{Swift}/BAT}
    \label{table: swift long}\\
    \midrule \midrule
  \endfirsthead

\bottomrule
\insertTableNotes
\endlastfoot

Time \tablenotemark{a}
& Flux \tablenotemark{b}
& Fluence Ratio \tablenotemark{c}
& $\eta$\tablenotemark{d} \\
(ks) & (Jy) & ($10^8$ Jy ms & ($10^{-11}$)\\
 & & erg$^{-1}$ cm$^2)$ &
\\
\midrule
$-20.6$ & $<$300-1000 & $<$3-8 & $<$130-300\\ 
$-19.6$ & $<$300-1000 & $<$3-8 & $<$120-300\\ 
$-18.6$ & $<$20-800 & $<$0.4-10 & $<$20-400\\ 
$-17.6$ & $<$4-70 & $<$0.02-0.9 & $<$0.6-40\\ 
$-16.6$ & $<$11-1000 & $<$1.0-13 & $<$40-500\\ 
$-15.6$ & $<$2-20 & $<$0.4-5 & $<$20-200\\ 
$-14.6$ & $<$3-900 & $<$0.11-40 & $<$5-2000\\ 
$-13.6$ & $<$400-1000 & $<$40-60 & $<$2000-2000\\ 
$-12.6$ & $<$500-2000 & $<$30-90 & $<$1300-4000\\ 
$-11.6$ & $<$120-2000 & $<$20-80 & $<$700-3000\\ 
$-10.6$ & $<$10-1000 & $<$0.8-20 & $<$30-1000\\ 
$-9.6$ & $<$800-2000 & $<$30-70 & $<$1000-3000\\ 
$-8.6$ & $<$900-2000 & $<$40-90 & $<$2000-4000\\ 
$-7.6$ & $<$14-2000 & $<$0.7-100 & $<$30-4000\\ 
$-6.6$ & $<$3-20 & $<$0.02-0.7 & $<$0.7-30\\ 
$-5.6$ & $<$2-10 & $<$0.08-0.5 & $<$3-20\\ 
$-4.6$ & $<$2-8 & $<$0.11-1.0 & $<$4-40\\ 
$-3.6$ & $<$4-60 & $<$0.2-5 & $<$9-200\\ 
$-2.6$ & $<$7-900 & $<$1-11 & $<$60-400\\ 
$-1.6$ & $<$500-2000 & $<$20-100 & $<$800-4000\\ 
$-0.6$ & $<$400-2000 & $<$20-90 & $<$900-3000\\ 
$0.4$ & $<$70-2000 & $<$10-70 & $<$400-3000\\ 
$1.4$ & $<$4-70 & $<$0.6-9 & $<$20-400\\ 
$2.4$ & $<$6-1000 & $<$0.8-13 & $<$30-500\\ 
$3.4$ & $<$5-600 & $<$0.2-8 & $<$9-300\\ 
$4.4$ & $<$5-900 & $<$0.04-40 & $<$1-1000\\ 
$5.4$ & $<$50-1000 & $<$0.3-20 & $<$14-900\\ 
$6.4$ & $<$130-1000 & $<$2-12 & $<$70-500\\ 
$7.4$ & $<$12-120 & $<$0.2-2 & $<$7-70\\ 
$8.4$ & $<$11-40 & $<$0.1-0.6 & $<$6-20\\ 
$9.4$ & $<$40-200 & $<$0.6-3 & $<$20-130\\ 
$10.4$ & $<$200-800 & $<$3-13 & $<$130-500\\ 
$11.4$ & $<$300-800 & $<$3-20 & $<$100-700\\ 
$12.4$ & $<$200-600 & $<$2-6 & $<$90-200\\ 
$13.4$ & $<$6-400 & $<$0.13-5 & $<$5-200\\ 
$14.4$ & $<$1-10 & $<$0.015-0.13 & $<$0.6-5\\ 
$15.4$ & $<$8-400 & $<$0.07-11 & $<$3-400\\ 
$16.4$ & $<$12-300 & $<$0.8-7 & $<$30-300\\ 
$17.4$ & $<$4-20 & $<$0.3-2 & $<$11-60\\ 
$18.4$ & $<$2-12 & $<$0.1-0.8 & $<$6-30\\ 
$19.4$ & $<$2-8 & $<$0.13-0.6 & $<$5-20\\ 
$20.4$ & $<$3-20 & $<$0.2-2 & $<$8-60\\ 
$21.4$ & $<$11-50 & $<$0.7-4 & $<$30-100\\ 
$22.4$ & $<$3-20 & $<$0.2-1 & $<$7-50\\ 
$23.4$ & $<$20-700 & $<$0.9-8 & $<$40-300\\ 
$24.4$ & $<$500-2000 & $<$2-6 & $<$100-200\\ 
$25.4$ & $<$500-1000 & $<$2-6 & $<$100-200\\ 
$26.4$ & $<$500-1000 & $<$3-6 & $<$110-200\\ 
$27.4$ & $<$120-2000 & $<$1-9 & $<$50-300\\ 
$28.4$ & $<$3-120 & $<$0.03-1 & $<$1-50\\ 
$29.4$ & $<$3-30 & $<$0.014-0.14 & $<$0.6-6\\ 
$30.4$ & $<$30-2000 & $<$0.15-5 & $<$6-200\\ 
$31.4$ & $<$800-2000 & $<$4-10 & $<$150-400\\ 
$32.4$ & $<$700-1000 & $<$3-7 & $<$130-300\\ 
$33.4$ & $<$700-1000 & $<$3-8 & $<$130-300\\ 
$34.4$ & $<$700-2000 & $<$4-9 & $<$150-400\\ 
$35.4$ & $<$900-2000 & $<$4-30 & $<$160-1300\\ 
$36.4$ & $<$60-1000 & $<$3-30 & $<$110-1300\\ 
$37.4$ & $<$7-1100 & $<$0.4-20 & $<$14-800\\ 
$38.4$ & $<$20-1300 & $<$0.3-20 & $<$11-800\\ 
$39.4$ & $<$20-1200 & $<$1-20 & $<$50-700\\ 
$40.4$ & $<$8-900 & $<$0.6-20 & $<$20-800\\ 
$41.4$ & $<$15-2000 & $<$0.11-30 & $<$5-1100\\ 
$42.4$ & $<$800-2000 & $<$11-30 & $<$400-1100\\ 
\end{longtable}   
\end{ThreePartTable}

\FloatBarrier

\begin{deluxetable}{c c c c}
\tablecaption{CHIME/FRB Upper Limits on Radio Emission from LGRBs detected by \textit{Fermi}/GBM \label{table: LGRBs fermi}
}
\startdata
\\
Time \tablenotemark{a}
& Flux \tablenotemark{b}
& Fluence Ratio \tablenotemark{c}
& $\eta$\tablenotemark{d} \\
(ks) & (Jy) & ($10^8$ Jy ms & ($10^{-11}$)\\
 & & erg$^{-1}$ cm$^2)$ &
\\ \hline
$-5.882$ & $<$2000-3000 & $<$2-7 & $<$90-300\\ 
$-4.882$ & $<$20-5000 & $<$0.3-5 & $<$12-200\\ 
$-3.882$ & $<$20-3000 & $<$0.3-8 & $<$11-300\\ 
$-2.882$ & $<$20-1100 & $<$0.013-0.9 & $<$0.5-40\\ 
$-1.882$ & $<$1100-5000 & $<$0.9-13 & $<$40-500\\ 
$-0.882$ & $<$2000-8000 & $<$3-7 & $<$130-300\\ 
$0.118$ & $<$2000-12000 & $<$3-30 & $<$130-1200\\ 
$1.118$ & $<$1400-4000 & $<$20-60 & $<$900-2000\\ 
$2.118$ & $<$1400-3000 & $<$20-40 & $<$900-1500\\ 
$40.118$ & $<$100-1100 & $<$0.006-0.06 & $<$0.2-2\\ 
$41.118$ & $<$700-2000 & $<$0.04-0.1 & $<$2-4\\ 
$42.118$ & $<$700-2000 & $<$0.04-0.09 & $<$2-4\\ 
$43.118$ & $<$200-3000 & $<$0.011-0.2 & $<$0.5-6\\ 
\enddata
\tablenotemark{a}{End time of a given ks bin for which the radio flux/fluence limit ranges apply. Times are relative to the detected high-energy emission, with negative times indicating emission prior to the high-energy emission. Start time for the range of fluxes/fluences is 1 ks prior to the end time.}

\tablenotemark{b}{Range of upper limits on the possible radio flux at the 99$\%$ confidence level for a 10-ms radio burst for the given ks time bin. The flux range is calculated using all \textit{LGRBs} detected by \textit{Fermi}/GBM within our sample.}

\tablenotemark{c}{Range of upper limits on the possible radio-to-high-energy fluence ratio at the 99$\%$ confidence level for a 10-ms radio burst for the given ks time bin. The fluence ratio range is calculated using all \textit{LGRBs} detected by \textit{Fermi}/GBM within our sample.}

\tablenotemark{d}{Range of upper limits on $\eta$ (unitless radio-to-high-energy fluence ratio assuming a 400-MHz radio emission bandwidth) at the 99$\%$ confidence level for a 10-ms radio burst for the given ks time bin. The fluence ratio range is calculated using all \textit{LGRBs} detected by \textit{Fermi}/GBM within our sample. The \textit{Fermi}/GBM high-energy range is 10 to 1000 keV, while the CHIME/FRB radio band is 400- to 800-MHz.}
\end{deluxetable}

\end{document}

%% file: auth.tex
\author[0000-0002-8376-1563]{Alice P.~Curtin}
  \affiliation{Department of Physics, McGill University, 3600 rue University, Montr\'eal, QC H3A 2T8, Canada}
  \affiliation{Trottier Space Institute, McGill University, 3550 rue University, Montr\'eal, QC H3A 2A7, Canada}
 \author[0000-0003-2548-2926]{Shriharsh P.~Tendulkar}
  \affiliation{Department of Astronomy and Astrophysics, Tata Institute of Fundamental Research, Mumbai, 400005, India}
  \affiliation{National Centre for Radio Astrophysics, Post Bag 3, Ganeshkhind, Pune, 411007, India}
\author[0000-0003-3059-6223]{Alexander Josephy}
  \affiliation{Department of Physics, McGill University, 3600 rue University, Montr\'eal, QC H3A    2T8, Canada}
  \affiliation{Trottier Space Institute, McGill University, 3550 rue University, Montr\'eal, QC H3A   2A7, Canada}
\author[0000-0002-3426-7606]{Pragya Chawla}
  \affiliation{Anton Pannekoek Institute for Astronomy, University of Amsterdam, Science Park 904, 1098 XH Amsterdam, The Netherlands}
\author[0000-0001-5908-3152]{Bridget Andersen}
  \affiliation{Department of Physics, McGill University, 3600 rue University, Montr\'eal, QC H3A 2T8, Canada}
  \affiliation{Trottier Space Institute, McGill University, 3550 rue University, Montr\'eal, QC H3A 2A7, Canada}
\author[0000-0001-9345-0307]{Victoria M.~Kaspi}
  \affiliation{Department of Physics, McGill University, 3600 rue University, Montr\'eal, QC H3A 2T8, Canada}
  \affiliation{Trottier Space Institute, McGill University, 3550 rue University, Montr\'eal, QC H3A 2A7, Canada}
\author[0000-0002-3615-3514]{Mohit Bhardwaj}
  \affiliation{Department of Physics, McGill University, 3600 rue University, Montr\'eal, QC H3A 2T8, Canada}
  \affiliation{Trottier Space Institute, McGill University, 3550 rue University, Montr\'eal, QC H3A 2A7, Canada}
\author[0000-0003-2047-5276]{Tomas Cassanelli}
  \affiliation{Department of Electrical Engineering, Universidad de Chile, Av. Tupper 2007, Santiago 8370451, Chile}
\author[0000-0001-6422-8125]{Amanda Cook}
  \affiliation{Dunlap Institute for Astronomy \& Astrophysics, University of Toronto, 50 St.~George Street, Toronto, ON M5S 3H4, Canada}
  \affiliation{David A.~Dunlap Department of Astronomy \& Astrophysics, University of Toronto, 50 St.~George Street, Toronto, ON M5S 3H4, Canada}
\author[0000-0003-4098-5222]{Fengqiu Adam Dong}
  \affiliation{Department of Physics and Astronomy, University of British Columbia, 6224 Agricultural Road, Vancouver, BC V6T 1Z1 Canada}
\author[0000-0001-8384-5049]{Emmanuel Fonseca}
  \affiliation{Department of Physics and Astronomy, West Virginia University, PO Box 6315, Morgantown, WV 26506, USA }
  \affiliation{Center for Gravitational Waves and Cosmology, West Virginia University, Chestnut Ridge Research Building, Morgantown, WV 26505, USA}
\author[0000-0002-3382-9558]{B.~M.~Gaensler}
  \affiliation{Dunlap Institute for Astronomy \& Astrophysics, University of Toronto, 50 St.~George Street, Toronto, ON M5S 3H4, Canada}
  \affiliation{David A.~Dunlap Department of Astronomy \& Astrophysics, University of Toronto, 50 St.~George Street, Toronto, ON M5S 3H4, Canada}
\author[0000-0003-4810-7803]{Jane F.~Kaczmarek}
  \affiliation{Dominion Radio Astrophysical Observatory, Herzberg Research Centre for Astronomy and Astrophysics, National Research Council Canada, PO Box 248, Penticton, BC V2A 6J9, Canada}
\author[0000-0003-2116-3573]{Adam E. Lanmnan}
  \affiliation{Department of Physics, McGill University, 3600 rue University, Montr\'eal, QC H3A 2T8, Canada}
  \affiliation{Trottier Space Institute, McGill University, 3550 rue University, Montr\'eal, QC H3A 2A7, Canada}
\author[0000-0002-4209-7408]{Calvin Leung}
  \affiliation{MIT Kavli Institute for Astrophysics and Space Research, Massachusetts Institute of Technology, 77 Massachusetts Ave, Cambridge, MA 02139, USA}
  \affiliation{Department of Physics, Massachusetts Institute of Technology, 77 Massachusetts Ave, Cambridge, MA 02139, USA}
\author[0000-0002-8912-0732]{Aaron B. Pearlman}
  \affiliation{Department of Physics, McGill University, 3600 rue University, Montr\'eal, QC H3A 2T8, Canada}
  \affiliation{Trottier Space Institute, McGill University, 3550 rue University, Montr\'eal, QC H3A 2A7, Canada}
\author[0000-0002-9822-8008]{Emily Petroff}
  \affiliation{Department of Physics, McGill University, 3600 rue University, Montr\'eal, QC H3A 2T8, Canada}
  \affiliation{Trottier Space Institute, McGill University, 3550 rue University, Montr\'eal, QC H3A 2A7, Canada}
  \affiliation{Anton Pannekoek Institute for Astronomy, University of Amsterdam, Science Park 904, 1098 XH Amsterdam, The Netherlands}
\author[0000-0002-4795-697X]{Ziggy Pleunis}
  \affiliation{Dunlap Institute for Astronomy \& Astrophysics, University of Toronto, 50 St.~George Street, Toronto, ON M5S 3H4, Canada}
\author[0000-0001-7694-6650]{Masoud Rafiei-Ravandi}
  \affiliation{Department of Physics, McGill University, 3600 rue University, Montr\'eal, QC H3A 2T8, Canada}
  \affiliation{Trottier Space Institute, McGill University, 3550 rue University, Montr\'eal, QC H3A 2A7, Canada}
\author[0000-0001-5799-9714]{Scott M.~Ransom}
  \affiliation{National Radio Astronomy Observatory, 520 Edgemont Rd, Charlottesville, VA 22903, USA}
\author[0000-0002-6823-2073]{Kaitlyn Shin}
  \affiliation{MIT Kavli Institute for Astrophysics and Space Research, Massachusetts Institute of Technology, 77 Massachusetts Ave, Cambridge, MA 02139, USA}
  \affiliation{Department of Physics, Massachusetts Institute of Technology, 77 Massachusetts Ave, Cambridge, MA 02139, USA}
\author[0000-0002-7374-7119]{Paul Scholz}
  \affiliation{Dunlap Institute for Astronomy \& Astrophysics, University of Toronto, 50 St.~George Street, Toronto, ON M5S 3H4, Canada}
\author[0000-0002-2088-3125]{Kendrick Smith}
  \affiliation{Perimeter Institute for Theoretical Physics, 31 Caroline Street N, Waterloo, ON N25 2YL, Canada}
\author[0000-0001-9784-8670]{Ingrid Stairs}
  \affiliation{Department of Physics and Astronomy, University of British Columbia, 6224 Agricultural Road, Vancouver, BC V6T 1Z1 Canada}
\newcommand{\allacks}{
A.P.C is a Vanier Canada Graduate Scholar.
SPT is a CIFAR Azrieli Global Scholar in the Gravity and Extreme Universe Program.
V.M.K. holds the Lorne Trottier Chair in Astrophysics \& Cosmology, a Distinguished James McGill Professorship, and receives support from an NSERC Discovery grant (RGPIN 228738-13), from an R. Howard Webster Foundation Fellowship from CIFAR, and from the FRQNT CRAQ.
A.B.P. is a Banting Fellow, a McGill Space Institute~(MSI) Fellow, and a Fonds de Recherche du Quebec -- Nature et Technologies~(FRQNT) postdoctoral fellow.
M.B. is supported by an FRQNT Doctoral Research Award.
C.L. was supported by the U.S. Department of Defense (DoD) through the National Defense Science \& Engineering Graduate Fellowship (NDSEG) Program. %
B.M.G. is supported by an NSERC Discovery Grant (RGPIN-2022-03163), and by the Canada Research Chairs (CRC) program. 
A.C. is supported by an Ontario Graduate Student Award.
Z.P. is a Dunlap Fellow.
The National Radio Astronomy Observatory is a facility of the National Science Foundation (NSF) operated under cooperative agreement by Associated Universities, Inc. S.M.R. is a CIFAR Fellow and is supported by the NSF Physics Frontiers Center awards 1430284 and 2020265.
F.A.D is supported by the U.B.C Four Year Fellowship.
E.P. acknowledges funding from an NWO Veni Fellowship.
P.S. is a Dunlap Fellow.
FRB Research at UBC is supported by an NSERC Discovery Grant and by the Canadian Institute for Advanced Research.
The Dunlap Institute is funded through an endowment established by the David Dunlap family and the University of Toronto.
}